\documentclass[preprint]{aastex}
\newcommand{\vi}{\hbox{$V\!-\!I$}}
\newcommand{\uv}{\hbox{$U\!-\!V$}}
\newcommand{\jk}{\hbox{$J\!-\!K$}}
\newcommand{\vk}{\hbox{$V\!-\!K$}}
\newcommand{\br}{\hbox{$B\!-\!R$}}
\newcommand{\bi}{\hbox{$B\!-\!I$}}
\newcommand{\ubvri}{\hbox{$U\!BV\!RI$}}
\newcommand{\ubv}{\hbox{$U\!BV$}}
\newcommand{\bvri}{\hbox{$BV\!RI$}}

\begin{document}

\title{M31 Globular Clusters: Colors and Metallicities\altaffilmark{1,2,3}}
\altaffiltext{1}{Work reported here based on observations made with the Multiple Mirror Telescope, 
a joint facility of the
Smithsonian Institution and the University of Arizona.}\altaffiltext{2}{Some of the data
presented herein were obtained at the W.M. Keck Observatory, which is operated as a scientific
partnership among the California Institute of Technology, the University of California, 
and the National Aeronautics and Space Administration. The Observatory was made possible
by the generous support of the W.M. Keck Foundation.}\altaffiltext{3}{This publication makes use of 
data products from the Two Micron All Sky Survey, which is a joint project of the University 
of Massachusetts and the Infrared Processing and Analysis Center, funded by the National 
Aeronautics and Space Administration and the National Science Foundation.}

\author{Pauline Barmby\\John P. Huchra}
\affil{Harvard-Smithsonian Center for Astrophysics, 60 Garden St., Cambridge, MA 02138} 
\email{pbarmby@cfa.harvard.edu, huchra@cfa.harvard.edu}
\author{Jean P. Brodie}
\affil{Lick Observatory, University of California, Santa Cruz, CA 95064}
\email{brodie@ucolick.org}
\author{Duncan A. Forbes}
\affil{School of Physics and Astronomy, University of Birmingham, Edgbaston, Birmingham B15 2TT, England, UK}
\email{forbes@star.sr.bham.ac.uk}
\author{Linda L. Schroder}
\affil{Lick Observatory, University of California, Santa Cruz, CA 95064}
\email{linda@ucolick.org}
\author{Carl J. Grillmair}
\affil{SIRTF Science Center, California Institute of Technology, Mail Stop 100-22, Pasadena, CA 91125}
\email{carl@ipac.caltech.edu}

\shorttitle{M31 Globular Clusters}
\shortauthors{Barmby et al.}

\begin{abstract}
We present a new catalog of photometric and spectroscopic
data on M31 globular clusters. The catalog includes
new optical and near-infrared photometry for a substantial 
fraction of the 435 clusters and cluster candidates.
We use these data to determine the reddening and intrinsic
colors of individual clusters, and find that 
the extinction laws in the Galaxy and M31 are not significantly different. 
There are significant (up to $0.2^m$ in \vk) offsets between the clusters' intrinsic
colors and simple stellar population colors predicted by 
population synthesis models; we suggest that these are due to
systematic errors in the models. 
The distributions of M31 clusters' metallicities and 
metallicity-sensitive colors are bimodal, with peaks 
at ${\rm [Fe/H]}\approx-1.4$ and $-0.6$.
The distribution of $V\!-\!I$ is often bimodal in 
elliptical galaxies' globular cluster systems, but is
not sensitive enough to metallicity to show bimodality
in M31 and Galactic cluster systems.
The radial distribution and kinematics of the two M31 
metallicity groups imply that they are analogs of the Galactic 
`halo' and `disk/bulge' cluster systems.
The globular clusters in M31 have a small radial metallicity 
gradient, suggesting 
that some dissipation occurred during the GCS formation.
The lack of correlation between cluster luminosity and metallicity
in M31 GCs shows that self-enrichment is not important in GC formation.

\end{abstract}

\keywords{galaxies: individual (M31) -- galaxies: star clusters -- 
  globular clusters: general}

\addtocounter{footnote}{-3}

\section{Introduction}

The study of the Andromeda Galaxy (M31) has been, and continues
to be, a keystone of extragalactic astronomy. The recognition
of M31 as an external galaxy by Hubble
marked the beginning of the field of extragalactic astronomy, and the recognition
of Populations I and II in M31 by Baade began the study of stellar 
populations. M31 is the nearest large galaxy to our own, and as
such has provided a wealth of information on stellar populations,
kinematics, and dynamics \citep[cf.][]{hod92}.
The results from the study of M31 provide an important benchmark
for comparison with more detailed results from the study of the Galaxy.

Globular clusters (GCs) are fossils of the earliest stages of galaxy 
formation. They are bright, easily-recognized packages comprising  
a stellar population with a single age and abundance. As such, their
integrated properties of location, abundance and kinematics
provide valuable clues to the nature and duration of galaxy 
formation. If galaxies form in a rapid monolithic dissipative 
collapse in which the enrichment timescale is shorter than the
collapse time, the halo stars and globular clusters should 
show a radial metallicity gradient. 
If galaxies are assembled from smaller, pre-enriched
fragments there should be no such gradient \citep{sz78}. 
Elliptical galaxies generally have more globular clusters
per unit luminosity (or unit mass) than spiral galaxies, and the
ellipticals' globular cluster systems (GCSs) are on average more metal-rich
\citep{har91}.
The GCSs of many elliptical galaxies show multi-modal metallicity
distributions, suggesting multiple star formation episodes or locations
\citep{za93}.
Some giant ellipticals, mostly in rich clusters, also have many more 
GCs than average for their luminosity. Several models have been 
proposed to account for these `extra' clusters: accretion of GCs
from smaller galaxies by capture or tidal stripping \citep{cmw98},
cluster formation in (spiral $+$ spiral $=$ elliptical) mergers \citep{az92,za93}, 
or a two-phase scenario where the metal-poor
clusters form (in the galaxy cluster potential) early in the galaxy collapse 
and the metal-rich clusters form later with the galaxy stars \citep{fbg97}.
The tidal stripping and two-phase models show better agreement with observations 
in that they predict that the metal-poor GCs should be more numerous than
the metal-rich ones, while the merger model predicts the reverse.
Multiple populations of GCs have also been identified in the Galaxy and
M31 (the only well-studied spiral galaxies GCSs); these have usually
been associated with the Population I and II stars, without 
further attempt to explain the GCS formation specifically.

M31's globular clusters were first recognized by \citet{hub32}.
The M31 globular cluster system has a unique
place in the study of globular cluster systems: it is the most
populous GCS in the Local Group, with about 800 proposed cluster candidates. 
Over 200 of these objects have been confirmed as clusters, 200 have been
shown not to be clusters, and the nature of the remaining objects
is unknown.  The study of the M31 GCS provides a bridge between
the study of the Galactic globular clusters, where most observations
are of individual stellar properties, and the study of most extragalactic 
clusters, where integrated properties are the only observables. 
In M31, most data are on integrated properties, but the advent of HST has
made individual stellar properties available 
in the form of color-magnitude diagrams \citep[e.g.][]{cmd1}, 
and ground-based adaptive optics systems will 
continue this trend. The same technological advances also 
extend the distance to which individual GCs can be observed:
for example, \citet{bau97} detect globular clusters around 
the Coma cluster galaxy IC~4051.
The M31 GCS will be important as a comparison in the study of the
integrated properties of these distant cluster systems.

Studies of the M31 globular cluster system are numerous: some
of the major attempts to catalog the system include \citet{vet62},
\citet{sar77}, \citet{bat80,bat87,bat93}
and \citet{cra85}. We attempt
to combine the existing catalogs of M31 GCs to make a comprehensive
catalog of confirmed clusters and good cluster candidates, and a
complete list of definitive {\it non-}clusters. We bring
together the published spectroscopic and photometric information
along with substantial amount of new photometry in the optical and
near-infrared, made possible by large-area mosaic CCD cameras
and IR array detectors. We determine 
the reddening for individual clusters, and use these data to 
examine the extinction in M31 as a whole and the intrinsic colors of the clusters.
We use the information to examine the
use of colors both to identify clusters and as metallicity 
indicators. This is an important issue for distant GCSs, where 
obtaining spectroscopic information is not feasible. 
The combination of spectroscopic and photometric information allows
us to search for multiple populations of globular clusters in M31
and determine those populations' characteristics.

\section{Catalog preparation}\label{catprep}
A study of the M31 cluster system requires a catalog that is
as complete and uncontaminated as possible, to avoid selection
biases and interlopers. Our catalog is based on the work of 
several previous authors, of which the catalog of \citet{bat87}
is the most comprehensive. To this we added the DAO catalog
\citep{cra85} and a list of cluster candidates near the nucleus by
\citet{bat93}. The \citet{bat87} and DAO
catalogs cover the entire galaxy in a fairly uniform manner.
To avoid introducing biases in the azimuthal distribution of clusters,  
we did not include the new cluster candidates of \citet{m98}
in our catalog since their fields cover only a small portion of the galaxy.
We pruned our catalog by removing objects
which the Bologna group classified as class `C', `D', or `E' (unlikely to
be clusters), unless they had been observed by another group.
We also compiled a complete list of candidates shown not to be 
clusters by high-resolution imaging or spectroscopy,
and removed these objects from our `cluster' catalog.

Naming the M31 globular clusters is complicated by the number of
works that have attempted to catalog the system.
The Bologna group's catalogs are the most extensive, so we retained
their numbering system. Following \citet*{hbk91} (hereafter HBK),
we added the number of the object in the 
``next most significant'' catalog to the Bologna numbers. 
These catalogs are the catalog of \citet{sar77} (indicated
without a letter after the dash), the `DAO' objects of \citet{cra85}
(indicated with a D after the dash), and the catalog of
\citet{vet62} (his Hubble or Baade numbers indicated
with H or B). Objects not in the Bologna catalog have numbers
beginning with `000-' and objects appearing only in their catalog 
have numbers ending in `-000'.
Of course, many objects appear in more than two catalogs, but
we refer the reader to the original papers \citep{sar77,bat87,vet62}
for further cross-identifications.
Note that the Bologna group maintained a separate numbering system
for their D-class objects, so 150D-000 is object \#150 in the
\citet{bat87} list of D-class objects and
279-D068 is \#279 in \citet{bat87} and \#68 in \citet{cra85}.

The finding charts in the Bologna group's papers were extremely
useful in correctly identifying the clusters and cluster candidates 
in the crowded M31 fields. However, we found several cases where the object
identified on the finding charts did not match the coordinates
in the table, considered relative to nearby objects. The 
coordinates in the \citet{bat87} table are the same as
those given in \citet{sar77}, so we take those
to be the correct positions. The objects incorrectly
shown on the finding charts and their correct positions are:
064-125 (about 1\arcmin\  east of indication), 208-259 (1\arcmin\ 
south and 20\arcsec\  east),
and 375-307 (15\arcsec\  east, 25\arcsec\  south).
The object identified on the finding chart as 375-307 is actually 
268D-D082.

The next step after constructing the object catalog 
was the construction of a photometry catalog; 
analyzing the color and color-metallicity distributions
and determining the reddening required that we 
compile as much photometric information as possible,
and that it be as accurate as possible.
For our catalog, we attempted to
find the ``best'' photometry for each object by searching
the literature, in the following order of
priority: (1) CCD photometry \citep{rhh92,rhh94,bat93,m98}, 
(2) photoelectric photometry (the series of
papers by Sharov, Lyutyi, and collaborators, some of which
are compilations of earlier photoelectric measurements), and (3)
photometry from photographic plates \citep{buo82,cra85,bat87}. 
We did not include the photographic $r$-band 
data of \citet{bat87} since these have a large zero-point
offset from the standard Cousins $R$-band \citep{rhh92}; we also
did not include photometric data marked as uncertain 
(although we did compare these data with our photometry) or with
given photometric errors $>0.1^m$.
There is no overlap between the CCD 
photometry datasets (they cover different parts of the galaxy),
so we did not need to choose between different observations
of the same object. For duplicate observations in the 
photoelectric data, we used the most recent ``average'' value,
which includes all of the observations by Sharov, Lyutyi,
and collaborators of a given object. For duplicate instances
of plate photometry, we used the most recent value.
Because the photoelectric and photographic data are in \ubv\ 
or \bv\ and the CCD data in subsets of \bvri, compiling data for
each object in as many filters as possible resulted in 
many of the objects having photometry from
multiple sources. Part of the motivation for our new
observations was to produce a set of consistent photometry
using the same identifications and aperture sizes for all objects.

In the near-infrared, there are fewer sources of photometry
\citep{fpc80,sit84,bo87,bo92,cm94}
and there is less overlap between them. We used the list of observations
in \citet{bo92}, which includes the earlier
IR papers, and added the data reported in \citet{cm94}.
Most observations of the same object by different groups
agreed very well, and we used the \citet{fpc80}
observations when these duplications occurred.
There were a few cases where duplicate observations did not 
agree (317-041, 029-090, 403-348, 373-305), and for these
we used the photometry with the smaller reported
error. 
Table~\ref{oldphot} contains the ``best'' photometry,
with references, for all of the objects in our catalog.
The comments section in this table indicates the existence of additional 
observational data not used in this study. These include high-resolution 
imaging (to confirm that objects are clusters), from \citet{rac91}
and \citet{rac92}; color-magnitude diagrams, from various 
authors; and high resolution spectroscopy, from \citet{dg97}
or \citet{djo97}.

Our resulting catalog has a total of 435 objects.
Prior to our new observations, all but 8 had at
least estimated $V$ magnitudes, and 330 had at least \ubv\ 
photometry. 158 had been observed spectroscopically, and 106
had infrared photometry. Our catalog
contains all the objects in the \citet{bat93} ``current best''
and ``extended'' samples, except for a few 
objects shown to be non-clusters after 1993.
About a dozen objects are possibly associated
with NGC~205. Because this galaxy is located well within the M31 globular 
cluster halo, in both position and radial velocity, 
it is not obvious how to determine which
clusters are actually associated with NGC~205 and which belong to the
M31 halo; different authors have come to different conclusions
on this subject \citep[cf.][]{dm88,rhh92}. We have flagged
these possible NGC~205 clusters in Table~\ref{oldphot}. We retain them
in our analysis of colors and color-metallicity relations, 
but omit them in studies of radial gradients and metallicity 
distribution.

We did not attempt to collect observational data for objects
declared to be non-clusters, but we did retain a list of the 
classifications (star, galaxy, H II region) of the objects and
the reference for this classification, given in Table~\ref{notclusters}. 

\section{Observations and Data Reduction}
\subsection{Optical photometry}

All of the new optical photometry reported here was collected using 
the 4-Shooter CCD mosaic camera \citep{fsh99} on the 1.2 meter telescope of the Fred L.
Whipple Observatory. Most of the observations were made in June 1998,
with additional data taken in October 1998 and January 1999.
The new data comprise 13 22-arcmin square fields in a grid centered on
M31, with a pixel scale of 0.67\arcsec\  per pixel.
Data reduction, beginning with the usual CCD processing steps
of bias subtraction and flat-fielding with dome flats, 
was performed in IRAF\footnote{IRAF is distributed by
the National Optical Astronomy Observatories, which are operated by the
Association of Universities for Research in Astronomy, Inc.,  
under cooperative agreement with the National Science Foundation.} 
using the {\sc mscred}, {\sc apphot}, and {\sc photcal} packages.

We performed photometric calibration of the M31 images 
using observations of \citet{lan} standard fields. We chose
positions for the fields to get standard stars on all four chips,
and also observed some smaller fields sequentially on all chips
in all five filters. We measured instrumental magnitudes of the
standard stars in large apertures using {\sc apphot}. To determine 
a photometric solution we fit
data from all four chips simultaneously, with separate zeropoints
and color terms for each chip, but only one airmass coefficient
for each color. For our June 1998 observing run we averaged
the color terms from the above procedure over all five nights,
and redid the photometric solutions. The airmass coefficients 
varied by $\lesssim0.02^m$ over the five nights (except in
$U$ where the photometric solution was poorer and the 
variation was $\sim 0.2^m$), and the zeropoint difference 
between chips was $\lesssim 0.10^m$. 
We expected a small zeropoint difference since the dome flats 
remove most, but not all, of the overall quantum efficiency differences
between the chips. The color term differences between chips
were on the order of $0.05^m$; again, these differences were
expected since the chips do not have exactly the same response curves.
A few of our fields also had deeper
observations taken in non-photometric conditions.  
We calibrated the photometry on the deeper images
by comparing stellar magnitudes to determine a mean magnitude
offset to the photometric images.

We identified the clusters and candidates by comparing our
images with the finding charts in the Bologna group's papers
\citep{bat80,bat87,bat93}.
Clusters not on the finding charts (DAO clusters, etc.) were located by 
offsetting from the nearest cluster marked on the chart.
Some clusters were difficult to identify, either because of
high local background or confusion between nearby objects.
These are marked with `ID' in the comments to Table~\ref{newphot}. 

We did simple aperture photometry of the clusters using 
{\sc apphot}; the results are in Table~\ref{newphot}. 
To match the aperture sizes of previous
photometry, we used an aperture of radius 12 pixels (8.0 arcsec)
for most clusters. For the few clusters near bright stars,
we measured the magnitude in a smaller aperture and
corrected it to the larger aperture size, using average
growth curves derived from other clusters in the
same field. These growth curves also showed that the 
choice of aperture size was reasonable, as $\sim94$\%
of the light from the clusters was contained within the
12 pixel radius.
Clusters with aperture-corrected magnitudes are marked in 
the comments to Table~\ref{newphot}.

Steep gradients in the galaxy light near the nucleus cause
two problems in aperture photometry of clusters: inaccurate centering 
of the cluster in the aperture, and systematic errors in the
background subtraction caused by steep gradients in the galaxy light.
For fields near the nucleus of the galaxy we performed
additional galaxy background removal. For each 
field, we subtracted an image of the smooth galaxy background
produced using a ring median filter \citep{sec95} and rescaled
the resulting image to have the same mode as the original
image. This produced very good galaxy subtraction to 
within $\sim  5$\arcmin\  of the nucleus. Comparison of photometry
on subtracted and non-subtracted images showed no changes
in photometric scale or zeropoint, but (as hoped) the photometric
errors were lower on the subtracted images because of 
lower sky background uncertainties.

Our 4-shooter fields in M31 overlap slightly. This
provides an opportunity to determine the precision of 
our photometry and photometric calibration by comparing photometry of
clusters which appear (always on different chips) in more 
than one field. The overlap regions are at the edges of the chips,
where accurate flat-fielding is more difficult and the 
photometric precision is slightly lower. 
The RMS differences between measurements of the same objects 
near the edges of different chips should therefore provide
an upper limit to our actual internal photometric uncertainties.
There are about 45 objects in the overlap regions; comparison of 
their magnitudes and colors shows that the scatter in the $V$ 
magnitudes is approximately $0.05^m$, and the
scatter in \bv, \vr,  and \vi\  colors is $\sim 0.08^m$.
Our $U$ observations were not deep enough to produce
reliable $U$ magnitudes for many clusters, so there
are not enough duplicate observations to determine
the scatter in \ub\ . However, the scatter in \ub\ 
is likely larger than that in the other colors and
we estimate it to be $\sim 0.15^m$.

\begin{figure}[t]
\includegraphics*[scale=0.9,angle=0]{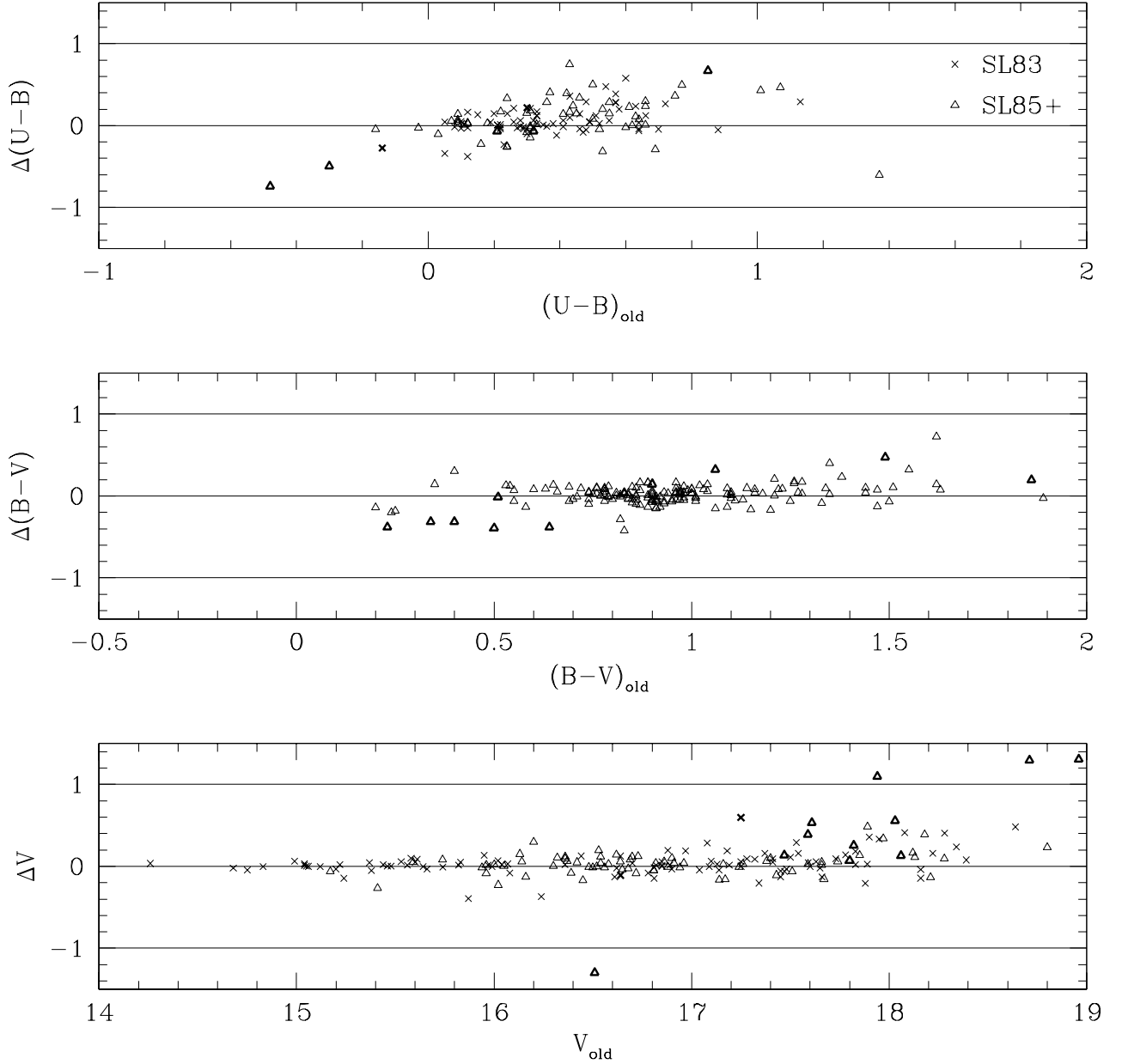}
\caption{Comparison of new photometry to previous photoelectric results.
Vertical axis in this and following figures is always (previous photometry)$-$(this work). 
SL83: \protect{\citet{sl83}}; SL85$+$: \protect{\citet{sl85}} and succeeding papers. 
Bold symbols are objects for which another photometric method agrees better with our results.
\label{pe_phot}}
\end{figure}

\begin{figure}[t]
\includegraphics*[scale=0.9,angle=0]{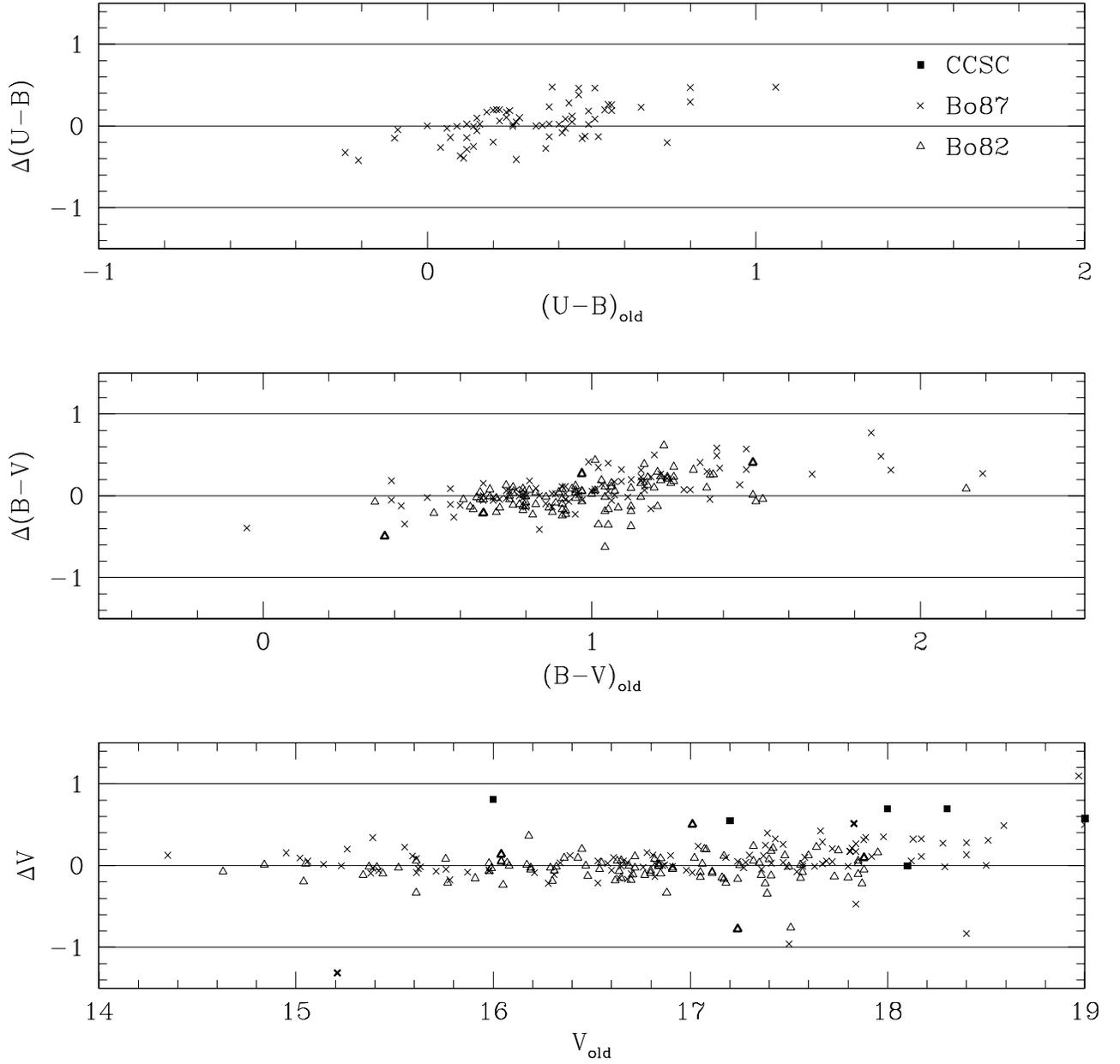}
\caption{Comparison of new photometry to previous photographic results.
CCSC: \protect{\citet{cra85}}; Bo87: \protect{\citet{bat87}}; Bo82: \protect{\citet{buo82}}.
Bold symbols are objects for which another photometric method agrees better with our results.
\label{pg_phot}}
\end{figure}

\begin{figure}[t]
\includegraphics*[scale=0.9,angle=0]{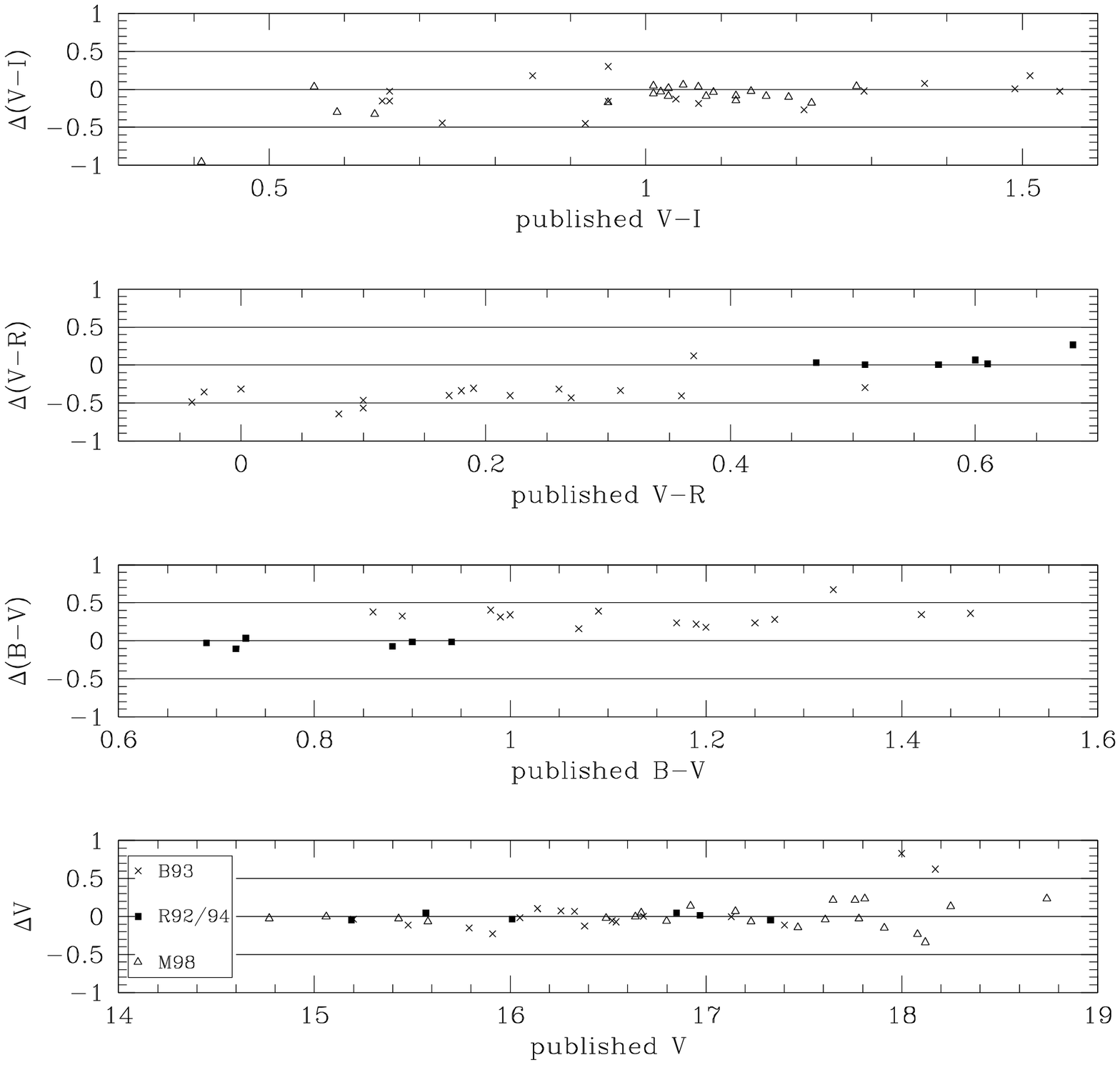}
\caption{Comparison of new photometry to previous CCD results.
B93: \protect{\citet{bat93}}; R92/94: \protect{\citet{rhh92,rhh94}}; M98: \protect{\citet{m98}}.
\label{ccd_phot}}
\end{figure}

We compared our optical photometry to published photometry,
separating the previous work by method (photoelectric, 
photographic, or CCD). Figures~\ref{pe_phot}, \ref{pg_phot},
and \ref{ccd_phot} and Table~\ref{optphotcomp} 
show the comparisons, in the sense (previous photometry$-$this work), 
for the various colors and filters. 
As expected, the scatter increases with $V$ magnitude,
but there is no evidence of a zeropoint offset or a
varying slope, with the exception noted below. 
In many objects where there is a 
large discrepancy between one type of published photometry
and our new work, our work agrees much better with 
one of the other types of published photometry.
These points are marked with bold symbols in the figures.
These large disagreements between the various photometry
sources are disappointing but unsurprising;
the last line of Table~\ref{optphotcomp} shows that, while
the photographic and photometric zeropoints agree well,
there is a large RMS scatter between the data sets.
Overall, we find that the published photometry is consistent
with our new data.

There was one published data set with which our results 
show marked disagreement - the CCD photometry of 
bulge clusters from \citet{bat93}.
Our \bv\ colors are bluer
and our \vr\ colors redder than theirs, in each case by
approximately 0.3$^m$. An obvious explanation, that our
$V$ magnitudes were 0.3$^m$ fainter, was not the case -- 
our $V$ magnitudes agree well with theirs. 
The \vr\ colors of \citeauthor{bat93} are bluer 
than those seen  for most globulars in M31 or the Galaxy, 
so we suspect that there may be systematic problem in
their photometry. Their \bv\ colors are redder than those of the 
average Galactic globular, but this is not unreasonable since these
clusters are near the M31 bulge and hence are more likely to be
metal-rich (and intrinsically red).
Other clusters in the same fields,
although further from the galaxy nucleus, show no
large offsets against previous photometry, and a check
of magnitudes on our ``raw'' and ``galaxy subtracted''
CCD frames shows that the galaxy subtraction procedure
did not substantially change the \bv\ colors.

There also appears to be a small offset in our \vi\ colors:
most of this is due to a few clusters with large
offsets, and the median offset is consistent with zero.
For the clusters where we disagree with \citet{m98}
they note that their $I$ magnitudes are suspect due to
nearby bad pixels. We have no explanation for the other large offsets,
except to note that our galaxy subtraction procedure
did not cause large changes in the \vi\ colors.
There is little comparison data for our \vr\ and \vi\ colors,
so we made a second inspection of the photometric solutions
in these colors. Our standard stars covered a wide range in colors,
and we found no bias in the residuals as a function of color,
so we are confident that any systematic errors affecting the 
$R$ and $I$ photometry are small.

\subsection{Near-infrared photometry}
Most of our new near-infrared data on the M31 clusters was taken
with the SAO IR camera \citep{tol98} on the 1.2m telescope
at FLWO, on 1998 Oct 27 and 28; conditions on both nights were
photometric. This two-channel camera has a 
5\arcmin\  field of view, which
required that we observe objects individually rather than
attempting to map the entire galaxy. We observed the objects without
published IR photometry in order of their $V$ magnitude, and collected
photometry for 122 new objects. 
We also obtained photometry of four objects from the 2MASS \citep{2m97}
scans of M31; these scans covered most of the galaxy but  
the short integration time of 6 seconds meant that only the brightest 
objects had acceptable signal-to-noise.

Our near-IR observations of objects and standard stars consisted
of 5 to 9 dithered frames in each of $J$ and $K$ per object.
Total integration times ranged from 140 to 240 seconds.
Data reduction for the IR data consisted of the
following steps: application of a non-linearity 
correction, dark subtraction,
flat-fielding, sky-subtraction, registration, and co-addition.
We made use of P. Hall's {\sc phiirs} package and a number
of our own IRAF scripts.

Flat fields were constructed by median-combining about
100 M31 cluster frames 
which did not include the galaxy nucleus.
Two sky-subtraction methods were used: for standard stars
and objects far from the nucleus we used running skies
(usually a median of the 8 frames nearest in time), and for
each object near the nucleus we observed a separate sky position,
median-combined those images to make a sky frame, and subtracted 
the sky from the object frames. We performed galaxy
subtraction on some of the co-added object frames, again using
the ring median filter.

We observed about a dozen \citet{eli82} standard stars per night, and
fit a two-component (zeropoint and airmass coefficient)
photometric solution using their measured aperture magnitudes.
We tried including a color term 
in the solution, but this did not improve the fit.
Others' experience also
indicates that the color term for this camera is negligible,
so we did not use it in our final solution. Residuals from the photometric
solution were $\lesssim 0.02^m$ for both nights and both filters.

We identified the clusters and candidates on the final coadded fields
by visual comparison with the optical finding charts. In addition 
to the target objects, many fields contained brighter 
objects with published photometry and/or additional, fainter objects. 
We measured aperture magnitudes for all the identified objects,
again using {\sc apphot}. We constructed growth curves for 
the brightest clusters and found that a 12-arcsec 
diameter aperture contained $\sim95$\% of the M31 clusters' light. 
This is comparable to the aperture sizes used in most previous 
IR photometry; the fact that a smaller aperture is required for
IR than for optical photometry is a consequence of the better 
seeing in the IR.

Following the procedure used for the optical photometry,
we performed both ``internal'' (night-to-night and
frame-to-frame) and ``external'' (previously published)
photometry comparisons. The within-night scatter is the standard deviation
of the differences in magnitudes of the same object
on different co-added images, and it is approximately $0.06^m$ in $J$
and $0.08^m$ in $K$. In both $J$ and $K$ this
scatter is $\sim0.02^m$ larger than the average photometric
errors, a measure of how ``photometric'' the conditions were. 
The scatter between observations of the same object on adjacent
nights is comparable to the within-night scatter.

\begin{figure}[t]
\includegraphics*[scale=0.9,angle=0]{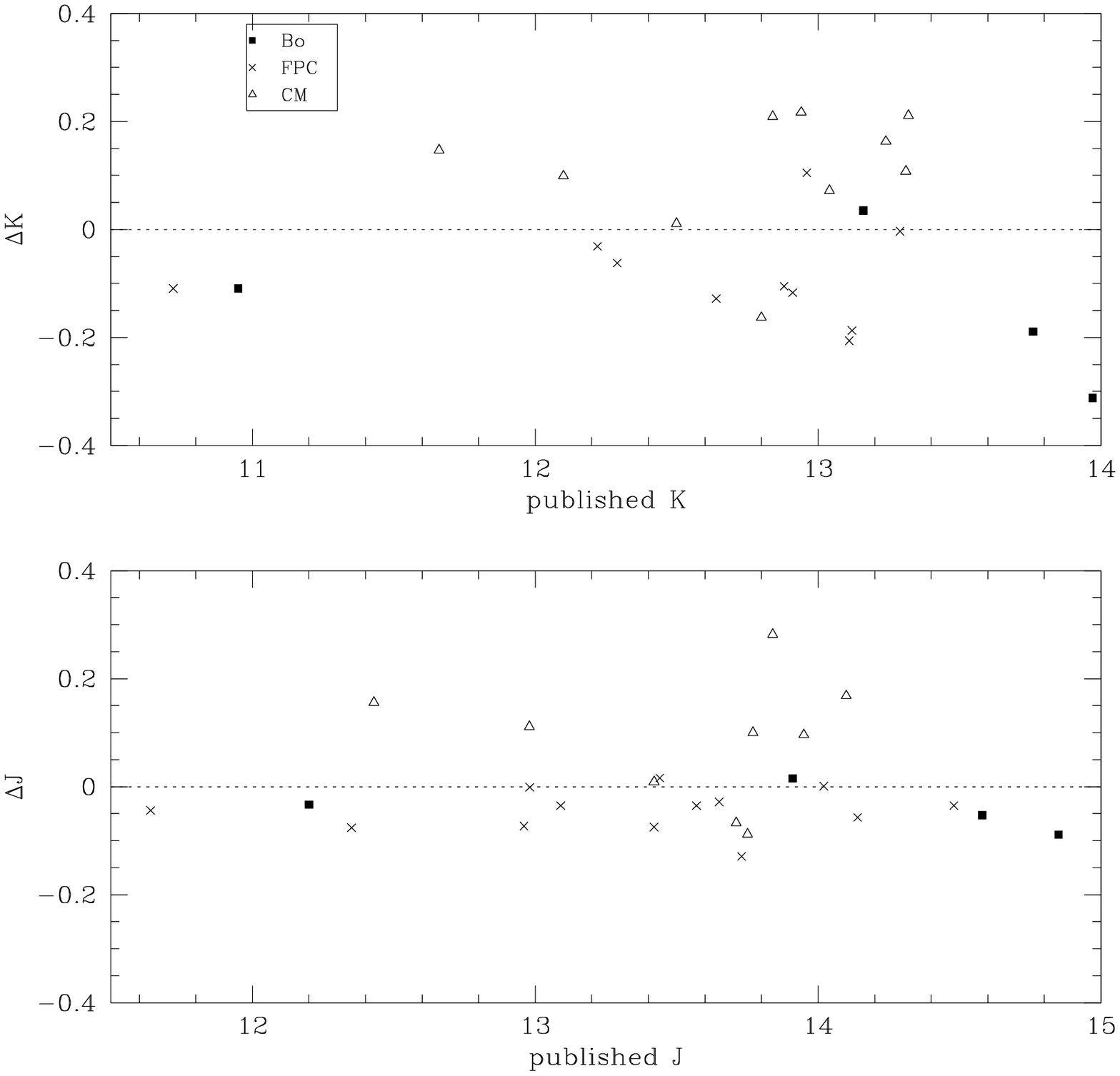}
\caption{Comparison of near-IR photometry to previous results.
Bo: \protect{\citet{bo87,bo92}}; FPC: \protect{\citet{fpc80}}; CM: \protect{\citet{cm94}}.
\label{ir_extphot}}
\end{figure}

Figure~\ref{ir_extphot} shows the results of the
external photometry comparison. Because the camera's field
of view is small compared to the size of M31, and because we
observed objects without previous photometry,
the number of comparison objects is small.
For the 24 objects with published photometry, both
offsets are consistent with zero: $\overline{\Delta K}=-0.014\pm0.031$,
$\overline{\Delta J}=+0.001\pm0.019$. The standard deviations 
of the photometry differences ($\sim0.15^m$ in $K$ and 
$\sim0.10^m$ in $J$) are larger than is comfortable,
but the small numbers make it difficult to tell if
this is due to some systematic problem in our own photometry or in
the previous work. 

\subsection{Spectroscopy}

We acquired new spectra of 61 cluster candidates, most with 
the Keck LRIS spectrograph \citep{oke95}
in 1995 December and 1996 September, and a few
with the MMT Blue Channel spectrograph in 1993 October.
With LRIS we used a 600 line/mm grating, giving 1.2 \AA/pixel 
dispersion from 3670-6200 \AA\ and a resolution of 4-5\AA.
With the Blue Channel, we used a 300 line/mm grating, giving 
3.2 \AA/pixel dispersion, spectral coverage from 3400-7200 \AA,
and 9-11 \AA\ resolution. Typical exposure times
were 4 minutes with LRIS and 15 minutes with the Blue Channel. 
We performed the usual reduction steps for CCD spectra (bias subtraction,
flat fielding, sky subtraction) using IRAF.
The wavelength calibration  used arc lamp spectra
taken in temporal proximity to the object spectra,
and the relative flux calibration used standard star
spectra taken on the same or adjacent nights; both were
also done in IRAF.

\begin{figure}[t]
\includegraphics*[scale=0.7,angle=90]{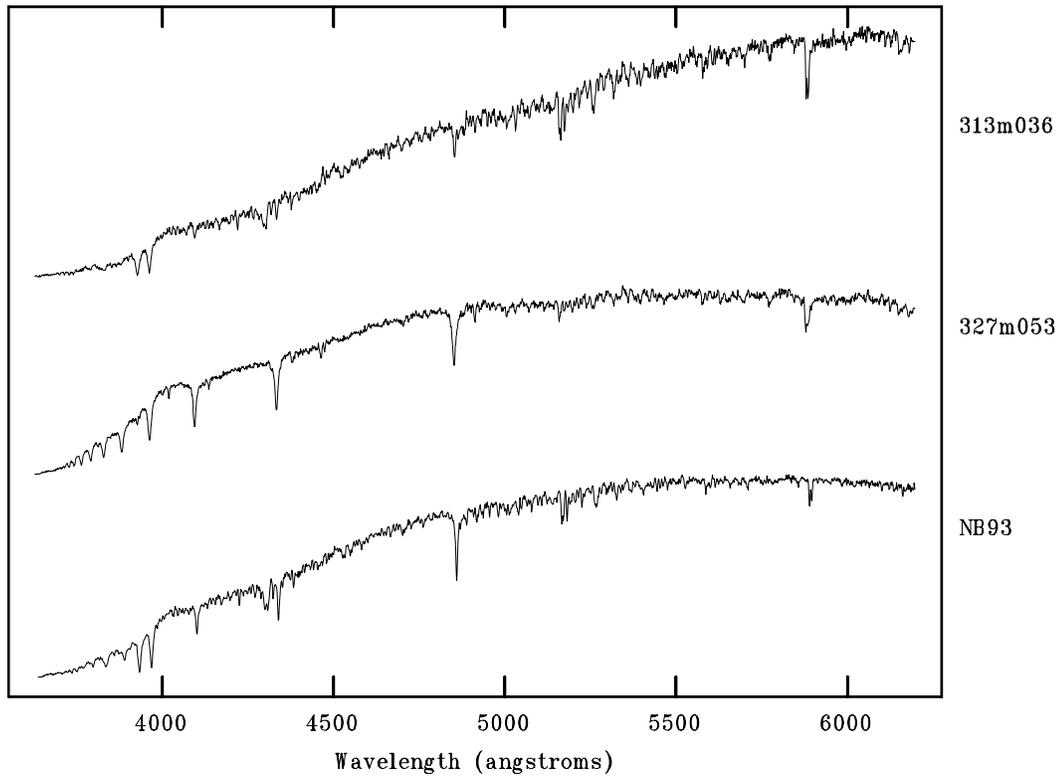}
\caption{Examples of new spectra, showing Galactic star (NB93), A-star like `young' globular 
(327-053), true globular (313-036).\label{newspec}}
\end{figure}

We used visual inspection of the spectra and radial velocity information to
determine which objects were bona fide globular clusters.
Objects with strong Na D lines, narrow line widths, continuum slope 
more appropriate to stars, and/or low radial velocities 
were classified as stars, while objects with large radial velocities
were classified as galaxies. Both classifications are noted in Table~\ref{newspectdata}.
We determined velocities of the clusters by cross-correlating their
spectra against spectra of template clusters with well-determined
velocities (225-280, 163-217, 158-213), taken on the same night,
using the {\sc xcsao} cross-correlation package.
The new velocities are in Table~\ref{newspectdata}.
Figure~\ref{newspec} shows examples of some of the new Keck spectra.

Several of the clusters with new spectra 
had spectra with large Balmer absorption
lines, but with cross-correlation velocities too 
large for them to be likely Galactic A stars.
Examining the archived spectra used by HBK, we identified several
more clusters with similar spectra, bringing the total number to 15. 
We tentatively classify these objects as young globular clusters;
other authors, including \citet{sar77} and \citet{ew88}, 
have similarly classified some of these objects.
We flag these objects as `young?' in Table~\ref{newspectdata}
and do not attempt to determine their metallicities. A detailed examination of
these objects will follow in a subsequent paper.

We modified the {\sc iraf} task {\sc sbands} to compute absorption
line indices according to the prescription of \citet{bh90}.
We tested this modified task on the archived MMT spectra of HBK
and found excellent agreement: our measurements differed from the
published values \citep{huc96} by less than 0.01$^m$, on average, for all 
indices.\footnote{When the archived spectra were transformed from
the original data format to {\sc fits} format, the details of
the original wavelength solution were lost, so we did not
expect to exactly duplicate the original index measurements.
One large discrepancy deserves mention: we found the
Fe5270 index for 034-096 to have a value of $0.0367\pm$0.013, but the
HBK value \citep[published in][]{egc4} is ten times as large.
We suspect that the HBK value is a typographical error, since
our value is more consistent with the other indices. The resulting
weighted metallicity is $-0.64\pm0.37$, compared to HBK's $0.31\pm2.08$.}
We measured the indices on flux-calibrated versions of our new spectra 
and determined the index errors using non-fluxed versions
of the same spectra, again according to the 
prescription given in \citet{bh90}. The measured
indices were combined using the metallicity calibration
defined in that paper to determine metallicities; the resulting
metallicity measurements are in Table~\ref{newspectdata}. 
It is clearly possible to do a more detailed metallicity analysis 
using the new Keck spectra, since they have better signal-to-noise
than most of the MMT spectra used by HBK; however,
since most of the spectroscopic data still come from that
paper, we used its methods to maintain consistency
across the cluster sample.

\subsection{Data summary}

We have compiled the results of our new photometry and spectroscopy 
with the existing data from the literature into a final catalog
of M31 cluster data. Optical photometry
is the only subset of the data where our new data significantly
overlaps with published work; to keep this data set as uniform as
possible, we used our photometry in preference to published
data unless our photometric errors were larger than $0.10^m$.
Of 435 clusters and cluster candidates, 268 have 
optical photometry in four or more filters, 224 have near-infrared 
photometry, 200 have velocities, and 188 have spectroscopic metallicities.
This catalog is the basis for the analysis to follow in the next section
and is, to our knowledge, the most comprehensive catalog of information 
available for M31 globular clusters and plausible cluster 
candidates. The catalog is available electronically at
\url{http://cfa-www.harvard.edu/~huchra/m31globs/}. The electronic
version contains additional information not given in this work, e.g.\
duplicate object names.

\section{Analysis}
Our classification of some M31 clusters as possibly young from their
spectra made us suspect that the cluster catalog could be contaminated
by other young objects. We checked this using \bv\ as a rough age indicator, 
since this color is available for the largest number of objects.
We found that most of the `young' clusters were
blue, with an average \bv\  of $0.37\pm0.07$; the average \bv\  
for all objects in the catalog was $0.87\pm0.02$. The two bluest Galactic globulars
in the June 1999 version of the \citet{h96} catalog 
have $(B\!-\!V)_0=0.40$ and 0.42, but 
these are the most- and least-reddened clusters in the catalog
(Terzan~5 has $E_{B-V}=2.37$ and NGC~7492 has $E_{B-V}=0.0$), so their colors
are somewhat suspect. The next bluest clusters have $(B\!-\!V)_0=0.55$.
It seems likely, then, that objects in M31 with $B\!-\!V<0.55$ are not
true globular clusters. There are 49 such objects in our catalog, and
their other colors are blue as well: for example, they lie along an
extension of the sequence in \bv\ vs. \ub\ formed by the redder objects.
We removed these blue objects and the remaining `young' clusters (which might
be reddened and thus have $B\!-\!V>0.55$) 
from our dataset before beginning the analysis.
\bv\ color is not a perfect selection criterion, of course: some objects do not have
\bv\ values, and some may have photometric errors which put them on
the wrong side of the boundary. However, except for a few 
clusters with poor-quality spectra, all of the spectroscopically-observed 
clusters with $B\!-\!V<0.55$ had already been identified as young from their spectra.
This suggests that our \bv\ criterion is reasonable.

\subsection{Reddening}\label{sec-red}
Previous photometric studies of the M31 GCS have dealt with the
problem of determining the cluster reddening in several ways.
\citet{fpc80} (hereafter FPC) corrected the colors of 35 clusters for reddening,
using the reddening-free parameter $Q_K$ from unpublished 
spectroscopic work by Searle. \citet{cra85} used the
intrinsic colors of the same 35 clusters to calibrate $(B\!-\!V)_0$ 
as a function of their spectroscopic slope parameter $S$.
Numerous authors \citep[e.g.][]{ir85,bg77,sh77} 
have used the globular clusters as reddening probes, often by
assuming them to have a single intrinsic color.
Most studies of the M31 GCLF \citep{rhh92,rhh94,sec92,kav97,gne97}
studied only clusters outside the ellipse used by \citet{rac91}
to define the outer boundary of the
M31 disk. These authors assumed that only foreground Galactic
extinction affected these `halo' clusters. 

For the disk clusters, there is almost certainly extinction
due to dust in the disk of M31, so merely correcting for Galactic extinction
is not sufficient. Determining the reddening from the total HI 
column density and dust-to-gas ratio is also not sufficient, since the clusters 
lie at different (and unknown) distances along the line of sight
through the M31 disk.
The assumption that the halo clusters suffer only foreground reddening
may also be incorrect: recent far-infrared observations
of spiral galaxies \citep{nel98,alt98}
indicate that the dust is more extended than the starlight.
It is important to test this by determining the reddening of the halo clusters.

With our larger database of multicolor photometry we attempted 
to determine the reddening for each individual cluster, using
correlations between optical and infrared colors and metallicity,
and by defining various `reddening-free' parameters.
To calibrate these methods we used the June 1999 version 
of the \citet{h96} database of Galactic GC parameters.
This database contains colors from \citet{pet93} and \citet{r86},
reddening values from multiple sources (mainly \citealt{rhs88},\citealt{w85}, and
\citealt{zi85}), and metallicities from multiple sources 
(mainly \citealt{zi85} and \citealt{az88}).
We corrected the colors for reddening using the $R_V=3.1$ 
extinction curve of \citet{ccm89}.
There are two major unavoidable assumptions in this procedure:
that the extinction law and the globular cluster intrinsic
colors are the same in the Galaxy and in M31. There is 
conflicting evidence on whether this first assumption is
correct: \citet{mas95} find $E_{U-B}/E_{B-V}=0.4-0.5$, 
\citet{ir85} found this ratio to be $1.01\pm 0.11$, 
and \citet{wk88} found it to be $0.6\pm0.2$.
Since there is no alternative optical-infrared extinction
curve for M31, using the Galactic curve is the only option.
We will show below that this is reasonable.

We performed linear regressions of intrinsic optical colors against
metallicity for the 88 Galactic clusters with $E_{B-V}<0.5$.
Colors used were $(B-V)_0,(B-R)_0,(B-I)_0,(U-B)_0,(U-V)_0,(U-R)_0,
(V-R)_0,(V-I)_0,(J-K)_0$, and $(V-K)_0$.
The correlation coefficients $r$ ranged from 0.91 for $(U-R)_0$ to    
0.77 for $(V-R)_0$; $(V-R)_0$ is a poor metallicity indicator because 
of its small range and we do not consider it further.
The color excess is determined from the observed
color, the metallicity-derived intrinsic color, and the
reddening ratio from \citet{ccm89}:
\begin{equation}\label{eq-colormet}
(X-Y)_0=a{\rm[Fe/H]} +b
\end{equation}
\begin{equation}\label{eq-redrat}
E_{B-V}=\frac{E_{B-V}}{E_{X-Y}}[(X-Y)-(X-Y)_0]
\end{equation}
We use $X-Y$ as generic notation to represent any color.

These color-metallicity relations allow us to check the 
assumption that the reddening laws in M31 and the Galaxy are the same.
To do this, we used the colors of the `old' M31 clusters with 
spectroscopic metallicities. For each cluster, we used the
above linear regressions to determine the color excess in
each color, then derived the various reddening ratios by dividing 
these color excesses by the color excess for \bv, also determined
from the intrinsic color-metallicity relation. 
Within the (admittedly large) uncertainties, the medians of these 
reddening ratios 
over all clusters were consistent with the Galactic values; see 
Table~\ref{extrat}. This result validates our use of the Galactic 
extinction curve to determine the reddening. We must still
show that the color-metallicity relations are the same for the
two sets of clusters and we do so in the following section.

To estimate the reddening for objects without spectroscopic
data, we also determined relationships between `reddening-free parameters'
and intrinsic colors. We derived all six possible reddening-free
parameters (hereafter referred to as Q-parameters) from the same Galactic
cluster \ubvri\ data used to calibrate the color-metallicity relations. 
The Q-parameters are defined as:
\begin{equation}
Q_{XYZ}\equiv(X-Y)-\frac{E_{X-Y}}{E_{Y-Z}}(Y-Z)=(X-Y)_0-\frac{E_{X-Y}}{E_{Y-Z}}(Y-Z)_0
\end{equation}
We then regressed these against the clusters' intrinsic
colors, and used the results to determine the color excess. Schematically:
\begin{equation}
Q_{XYZ} \Rightarrow (X-Y)_0 \Rightarrow E_{B-V}
\end{equation}
The correlation coefficients for the Q-parameters were poorer than those
for the color-metallicity relations, ranging from 0.80 for               
$Q_{BVR}$ to 0.27 for $Q_{VRI}$.
Since there is significant scatter in all of these correlations
(due to age or `second parameter' effects?)
applying them will yield only a rough estimate of the
individual cluster reddening, and we do not attempt to
treat the results in a statistically rigorous manner.

\begin{figure}[t]
\includegraphics*[scale=0.9,angle=0]{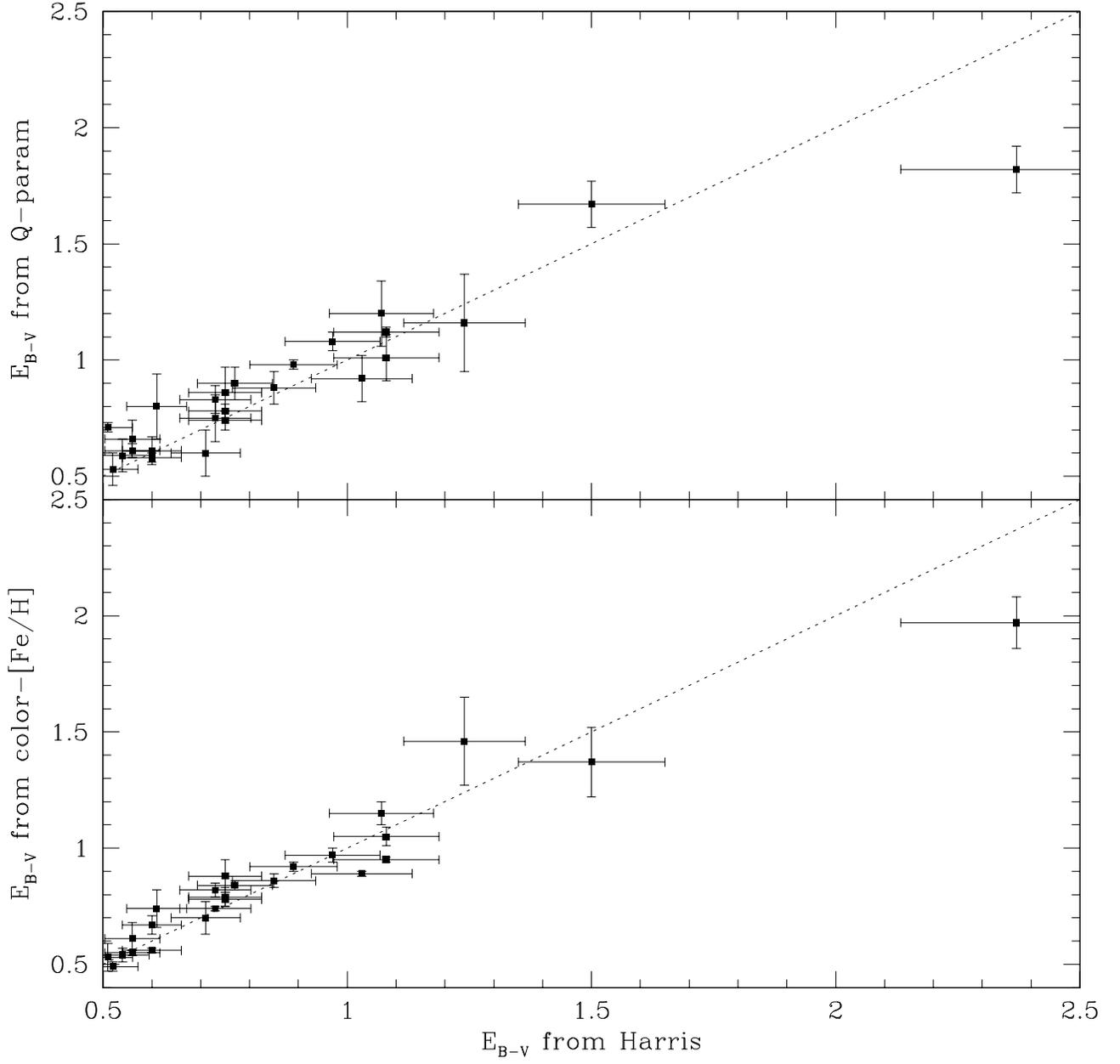}
\caption{High-reddening Galactic clusters: $E_{B-V}$ from \protect{\citet{h96}} 
vs. $E_{B-V}$ from new methods.\label{galred}}
\end{figure}

Our final reddening determination used seven of the nine colors
in Table~\ref{extrat} (we dropped \ub\ and \jk, since these colors
are not very sensitive to reddening) and all six Q-parameters.
For each of the two methods we averaged the results 
over all colors or parameters to produce one
value of $E_{B-V}$ per method. The standard deviations of
these averages serve as an estimate of the precision of the
methods.  
We tested the methods first on 25 heavily-reddened Galactic 
clusters not used for the calibration. The results were 
encouraging -- the precision of both methods, defined as 
$\sigma_{E_{B-V}}/\overline{E_{B-V}}$, had a median value of $\sim7$\%. 
The two methods agreed quite well both with each other and
with the color excesses from the Harris catalog: the average
offset between $E_{B-V}$ from the Q-parameter method and the Harris value
was $0.03\pm0.03$; for the color-metallicity method the average 
offset was $0.00\pm0.02$ (see Figure~\ref{galred}).

To determine the reddening for the M31 clusters we combined
the results from the two methods, subtracting 0.03 from the Q-method
results because of the offset noted in the previous paragraph.
We examined the errors in the cluster reddenings from the 
two methods, and determined that the errors were the same
when the error in [Fe/H] was approximately 0.4, and that the 
error in the metallicity-derived $E_{B-V}$ increased 
dramatically for ${\sigma}_{\rm [Fe/H]}>0.7$.  

The $E_{B-V}$ and [Fe/H] errors are related by
\begin{equation}
{\sigma}_{E(B-V)} = \left| \frac{\partial \overline{E_{B-V}}}{\partial{\rm [Fe/H]}} \right| {\sigma}_{\rm [Fe/H]}
= \left( \frac{1}{N}\sum \frac{E_{B-V}}{E_{X-Y}} a_{X-Y} \right) {\sigma}_{\rm [Fe/H]}
\end{equation}
(where $N$ is the number of colors and $a$ is the same as in equation~\ref{eq-redrat}). 
We weighted the metallicity-derived $E_{B-V}$ by the inverse of the
bracketed term in the above equation.
The zeropoint of the weighting was set so that the weight
of $E_{B-V}({\rm [Fe/H]})$ would be zero at ${\sigma}_{\rm [Fe/H]}=0.7$, 
and the weight for $E_{B-V}({\rm Q})$ was set to give the two
methods equal weight at ${\sigma}_{\rm [Fe/H]}=0.4$.  

To check our results, we compared our reddenings 
with the $E_{V-K}$ given for 34 clusters in FPC. 
A typical error in our values of $E_{B-V}$ for these clusters was 0.04.
For 24 of these clusters FPC quote $E_{V-K}=0.28$, which corresponds
to $E_{B-V}=0.10$, their value for the foreground reddening. For these
clusters our $E_{B-V}$ ranges from 0.01 to 0.18 with a mean  
$E_{B-V}=0.13\pm0.02$. For the 10 clusters with larger reddening, 
we find the median $E_{B-V}/E_{V-K}=2.7\pm0.2$, which is consistent
with the \citeauthor{ccm89} value of 2.75, or the FPC value of 2.8. 

We also compared our results with the predicted $E_{B-V}$ from
the Galactic dust maps of \citet{sfd98}, hereafter SFD. 
Since their map does not
account for reddening internal to the M31 disk, we compared only
reddening for objects in the halo, as defined by \citet{rac91}.
We were able to determine a reddening for 60 of these clusters, 
with typical errors of 0.06 in $E_{B-V}$.
The mean offset (SFD$-$our value) is $-0.02\pm0.01$, consistent 
with zero. However, the standard deviation of the offset (0.08) is large, 
and, for low reddening, our values scatter between 0 and 0.2 and  
show little correlation with the SFD results (which
have values between 0.05 and 0.1).  
The SFD maps show a large reddening for one cluster (462-000), and 
for it the agreement is fairly good: the SFD maps
give $E_{B-V}=0.29$ and we find $E_{B-V}=0.28\pm0.16$.
From the comparisons with SFD and FPC, we estimate that our 
values of $E_{B-V}$ 
have total errors between 0.05 and 0.10. These are large 
errors, but we believe this method is preferable to the
alternatives of correcting only for the foreground reddening or
doing no reddening correction at all.

\begin{figure}[t]
\includegraphics*[scale=0.9,angle=0]{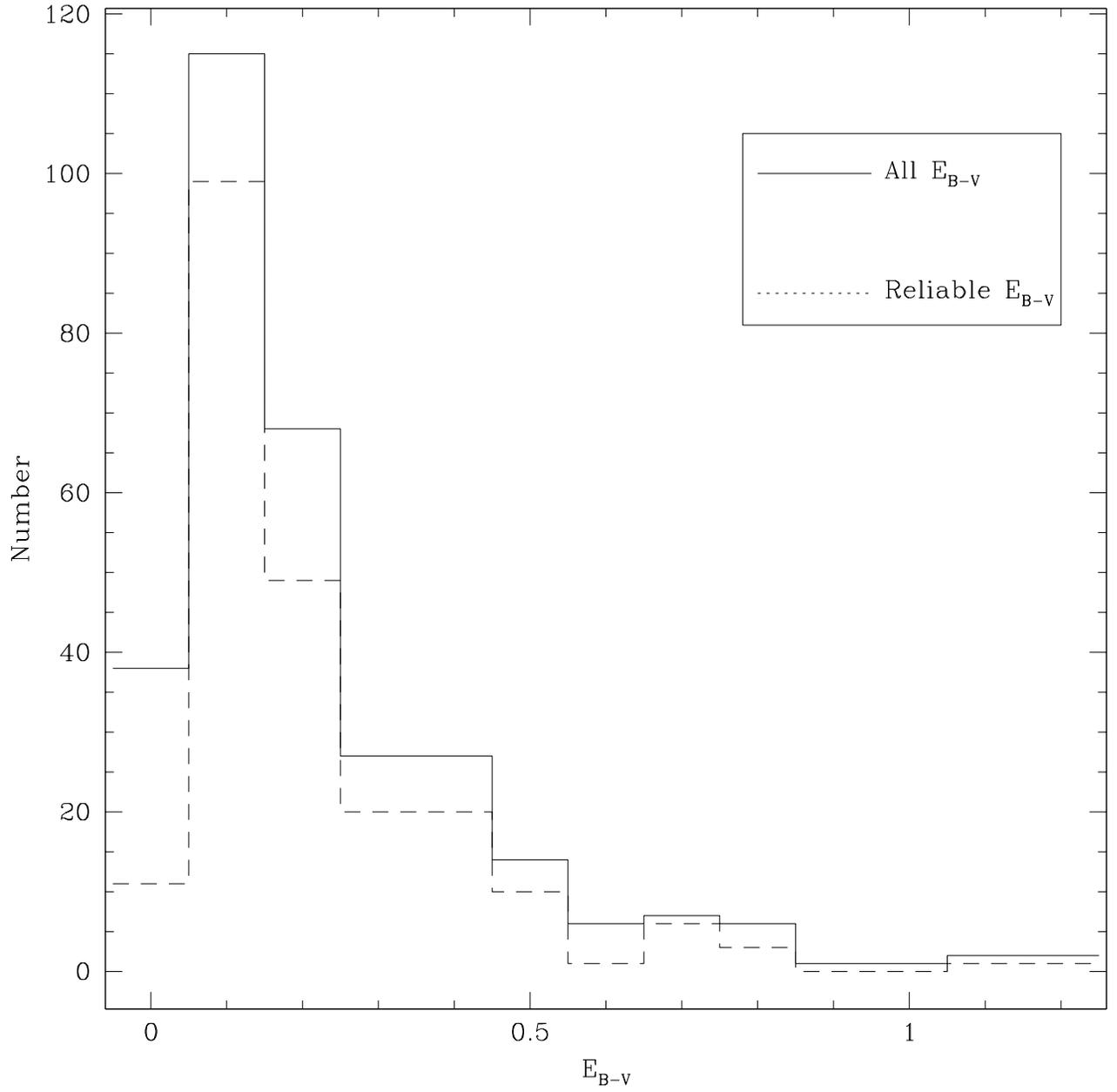}
\caption{Distribution of $E_{B-V}$ for M31 clusters\label{ebvdist}}
\end{figure}

We estimated reddenings for all the M31 clusters with sufficient data,
a total of 314 objects. Some values are more reliable than others:   
unreliable measurements are those with with no reddening errors
(those with only one color-metallicity relation
or Q-parameter) and those with large reddening errors
(arbitrarily chosen as 
$\sigma_{E_{B-V}}/\overline{E_{B-V}}> 0.5$ for $\overline{E_{B-V}}> 0.15$,
$\sigma_{E_{B-V}}/\overline{E_{B-V}}> 1.0$ for $\overline{E_{B-V}}< 0.15$).
We do not use these reddenings in the analysis that follows.
The distribution of the 221 reliable reddenings (Figure~\ref{ebvdist})  
has a mean of $E_{B-V}=0.22$, a median of $E_{B-V}=0.16$, and a standard
deviation of 0.19. Three-quarters of the clusters have $E_{B-V}<0.27$.
The largest `reliable' reddening is that of 037-B327, $E_{B-V}=1.38$;
\citet{vdb69} notes that this cluster is ``the most highly-reddened 
cluster known in M31''. If this reddening value is correct, 037-B327 
is twice as luminous as 000-001 ($=$G1), one of the brightest M31 GCs. 
Assuming an M31 distance modulus of 24.47 \citep{sg98,hol98} means that 
037-B327 has $M_V\approx-12$ and is more than four times 
as luminous as the brightest Galactic GC \citep[$\omega$~Cen at $M_V=-10.29$;][]{h96}.
This object is puzzling: as \citeauthor{vdb69} states, there is no obvious reason
why the intrinsically brightest GC in M31 should also be the most heavily-reddened.
However, the nature of 037-B327 is still uncertain:
the reddening estimate is from color information 
alone, and a spectrum of this object would be extremely valuable.

\begin{figure}[t]
\includegraphics*[scale=0.9,angle=0]{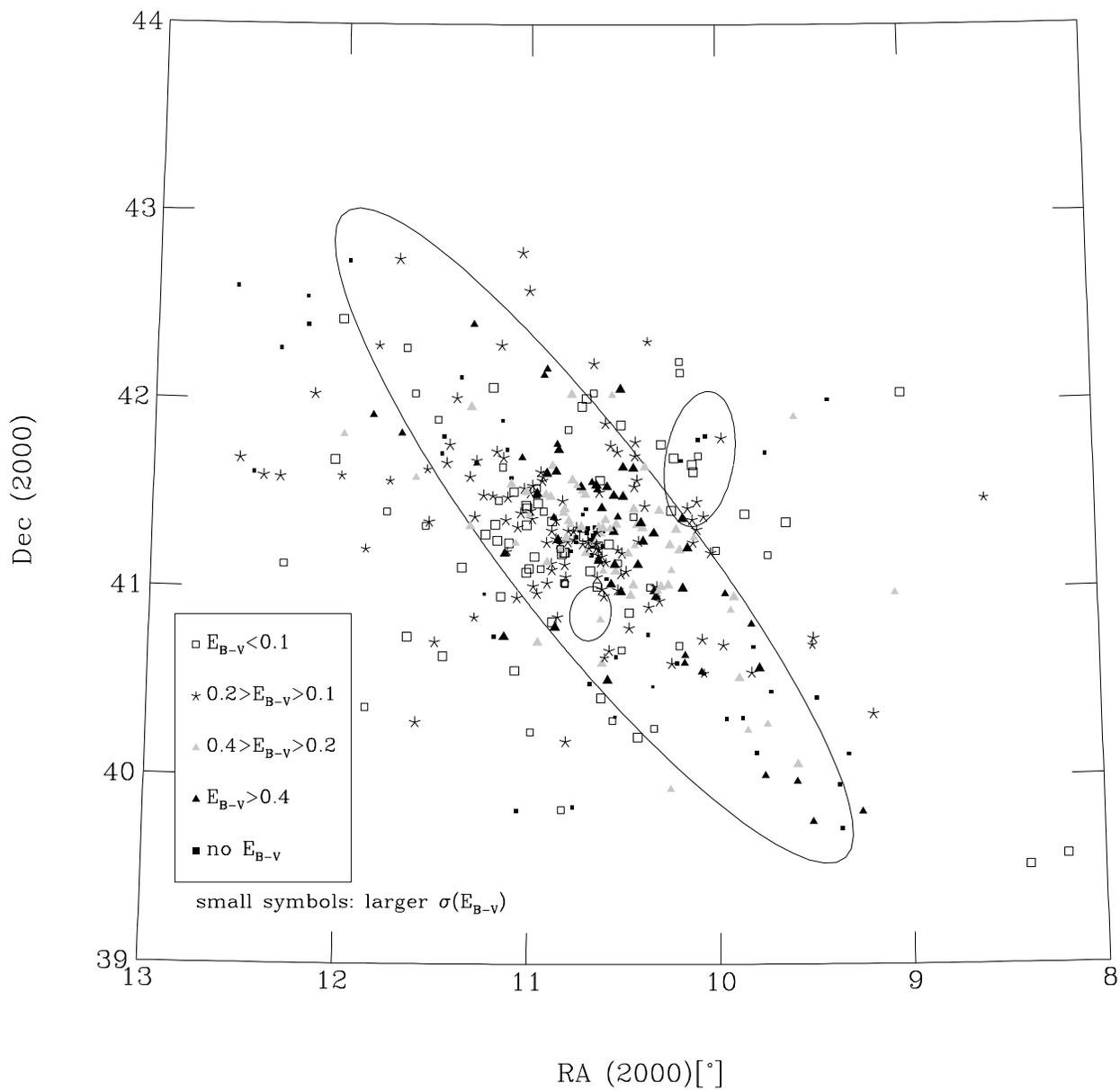}
\caption{Map of M31 globular clusters in RA and Dec, in groups according to reddening.
Large ellipse is M31 disk/halo boundary as defined by Racine (1991); smaller
ellipses are $D_{25}$ isophotes of NGC~205 (NW) and M32 (SE).\label{redmap}}
\end{figure}

All of the reddening values are shown in as functions of 
position in Figure~\ref{redmap}.
The maps appear reasonable in that the objects with the lowest reddening
are spread across the disk and halo, while those with the highest
reddening are concentrated in the galaxy disk.
The higher-reddening clusters in the disk tend to lie on the northwest 
side of the major axis. This accords with statements in previous work
\citep{ir85,ew88} that this side of the disk is nearer to us along the line of sight.
We note that a substantial number of clusters outside the `halo' 
boundary have $E_{B-V}>0.1$. While some of these values are
undoubtedly due to the large errors in our method, some 
clusters (such as 004-050, with $E_{B-V}=0.19\pm0.04$) have
reddening values that are very consistent over a number of colors
and Q-parameters. It seems unlikely that some systematic problem
in our method or photometry could affect all of the individual
colors to make them give the same erroneous reddening.
Two possibilities remain: (1) the M31 dust distribution extends
to greater projected distances than previously suspected, and/or (2)
the Galactic foreground extinction in the direction of M31 is
patchy on scales smaller than the SFD spatial resolution of 6.1\arcmin.
In either case, the assumption that the M31 halo clusters are subject to only 
foreground reddening is in some doubt and should be re-examined.

\subsection{Color-metallicity relation}\label{sec-colormet}
We showed in Section~\ref{sec-red} that the assumption 
of a similar reddening law in M31 and the Galaxy was reasonable.
The second major assumption made in our reddening correction
procedure was that the relation between intrinsic color and
metallicity is the same for M31 GCs and Galactic GCs. We
tested this assumption by doing BCES bisector linear fits
\citep[as described in][]{ab96} of color against 
metallicity for the M31 and Galactic clusters and comparing
the results. The bisector fit is appropriate since we are
interested in both the case where metallicity is used to predict color,
as in the determination of reddening, and the case where 
color is used to predict metallicity, as follows in Section~\ref{sec-metdist}. 
(To do the actual predictions we used the BCES(Y$|$X) fit,
an extension of the ordinary least-squares fit which allows for measurement
error in both variables and intrinsic scatter.  For reference, these
fits for the Galactic data are given in Table~\ref{color-met}.)

For the Galactic color-metallicity fits we used the same data used
to determine the color-metallicity relations in Section~\ref{sec-red};
we estimated the intrinsic color errors as:
\begin{equation}\label{eq-colorerr}
{\sigma}_{X-Y}=({{\sigma}_{\rm phot}}^2 + ({\sigma}_{E(B-V)}E_{X-Y}/E_{B-V})^2)^{1/2}
\end{equation}
We set ${{\sigma}_{\rm phot}}$ to $0.02^m$, as this is a typical
uncertainty in \citet{r86}, one of the main sources of integrated colors, 
and ${\sigma}_{E(B-V)}$ as $0.1E_{B-V}$, following \citet{h96}.
We set $\sigma_{\rm [Fe/H]}$ for the Galactic clusters to 0.10 dex;
typical uncertainties in \citet{zi85} \citep[one of the major sources for][]{h96}
are 0.15 dex, but many of the [Fe/H] values are averages from several sources.
For the M31 fits we used 101 M31 clusters with reliable reddening and 
$\sigma_{\rm [Fe/H]}<0.5$ dex. We estimated color errors using equation~\ref{eq-colorerr},
with ${{\sigma}_{\rm phot}}$ set to $0.04^m$ and ${\sigma}_{E(B-V)}$
to our measured value. 

The BCES method produces estimates of the uncertainty in the slopes
and intercepts of the linear fits, so one way to compare the two sets
of fits is to compare the ratio of the parameter differences to the
parameter uncertainty. However, unlike the case for ordinary least-squares fitting,
the distribution of this ratio in the case of the null hypothesis
is unknown, so it is impossible to determine its statistical significance.
A more empirical approach is to simply compare the predictions of the
two fits. We determined the
differences in color predicted by the two fits at the metal-rich (red) 
and metal-poor (blue) ends of the data range,
and compared these to the rms color residuals of the fits.
We found the differences between the Galactic and M31 fits to be comparable
to the fit residuals for all the colors.

\begin{figure}[t]
\includegraphics*[scale=0.9,angle=0]{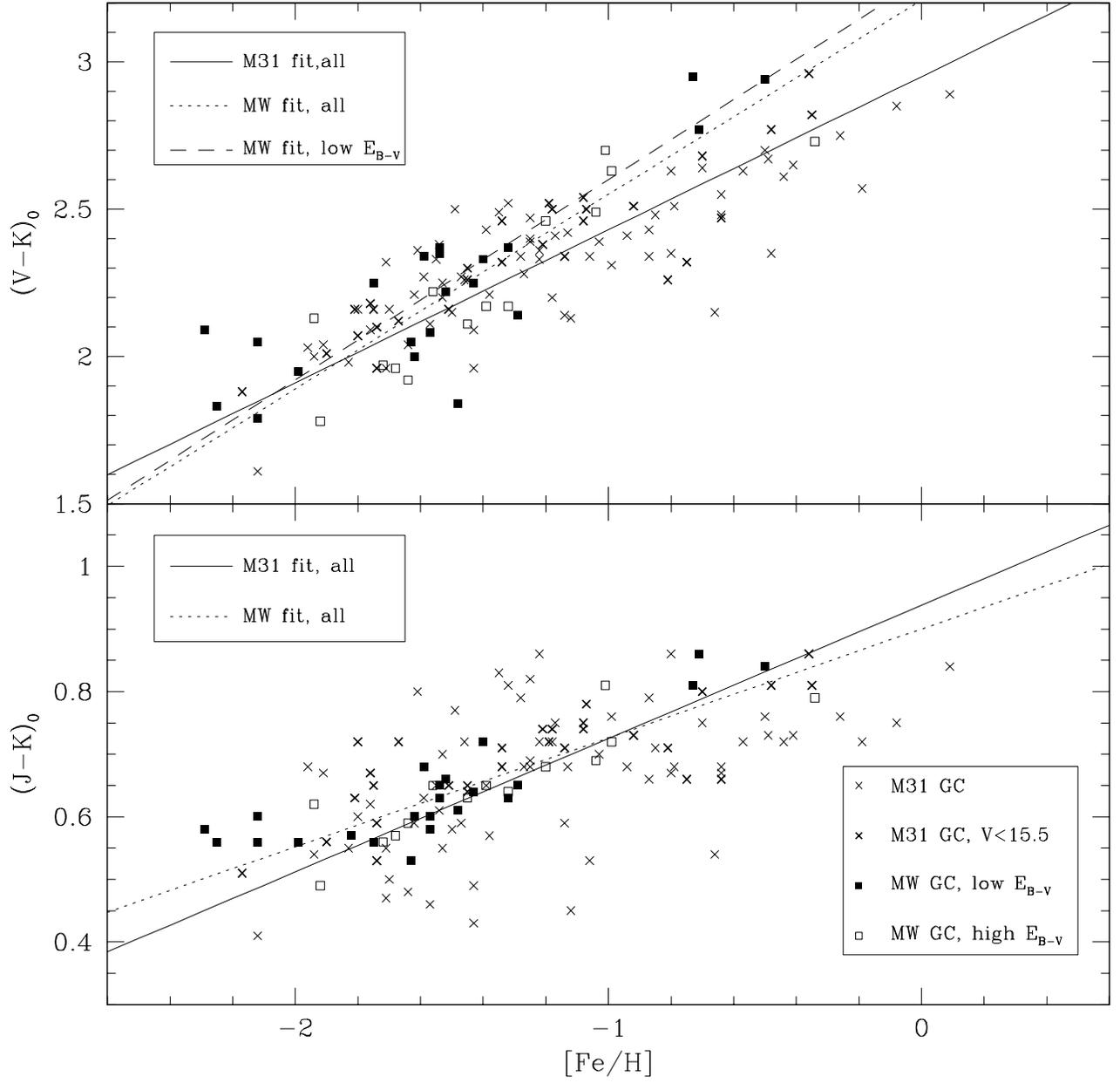}
\caption{$(J-K)_0$ and $(V-K)_0$ color-metallicity relations for M31 and 
Galactic GCs.\label{ir_colormet}}
\end{figure}

The $(V\!-\!K)_0$ and $(J\!-\!K)_0$ fits are shown in Figure~\ref{ir_colormet};
these deserve particular attention for several reasons.
The Galactic relations as originally derived by \citet{bh90}
rely on only 23 low-reddening calibrators
in their Table~5A.\footnote{The text of \citet{bh90}
indicates that their Table~5 contains the `raw' (i.e.\ uncorrected
for reddening) colors. This is incorrect -- {\it the colors in this table
have already been reddening-corrected.}} Only four of these Galactic
calibrators have ${\rm [Fe/H]}>-1.2$. To maintain consistency with the optical
color-metallicity fits, we did the optical-infrared color-metallicity fits
using the metallicity and reddening values in \citet{h96}, rather than 
those in \citeauthor{bh90}'s table. We also added the 14 ``high-reddening''
clusters in their Table~5B to see whether this made a difference to the
fits. As Figure~\ref{ir_colormet} shows, adding the additional clusters
made a difference for $(V\!-\!K)_0$, bringing the fit closer to that
for the M31 clusters (the second $(V\!-\!K)_0$ row in Table~\ref{color-met}
shows the Galactic fit with all clusters included). Calibrating
the color-metallicity relation with a small number of clusters 
means that even a small change in the input data can change the result.

For $(V\!-\!K)_0$ there are several M31 clusters which 
are either too blue for their metallicities or too metal-rich
for their colors, compared to the Galactic clusters
and the bulk of the M31 clusters. We have examined the spectra
of these clusters and their photometry, and find no obvious problems
with either. These clusters are not different from the bulk of 
M31 clusters in any obvious way (location, reddening, H$\beta$
strength, etc.), and we are unable to explain their anomalous
colors.

There is little difference in the $(J\!-\!K)_0$ fit when all clusters,
instead of just low-reddening ones, are included.
This is unsurprising since \jk\ is much less sensitive to reddening
than \vk. However, the $(J\!-\!K)_0$ vs. metallicity fit shows much 
larger rms residuals for the M31 clusters than the Galactic clusters. 
If we restrict the M31 data to the 31 clusters brighter than 
$V=15.5$ -- presumably these should have smaller photometric and
spectroscopic errors because they are brighter -- the
residuals are much closer to the Galactic values, although the
fit does not change significantly.
This suggests that errors in the \jk\ photometry may have been
underestimated, and points to the need for precise \jk\ colors
if $(J\!-\!K)_0$ is to be used as a metallicity indicator.
As Table~\ref{color-met} shows, $(J\!-\!K)_0$ is not as sensitive
to metallicity as most of the other colors; its advantage as a
metallicity indicator is its insensitivity to reddening.

\begin{figure}[t]
\includegraphics*[scale=0.9,angle=0]{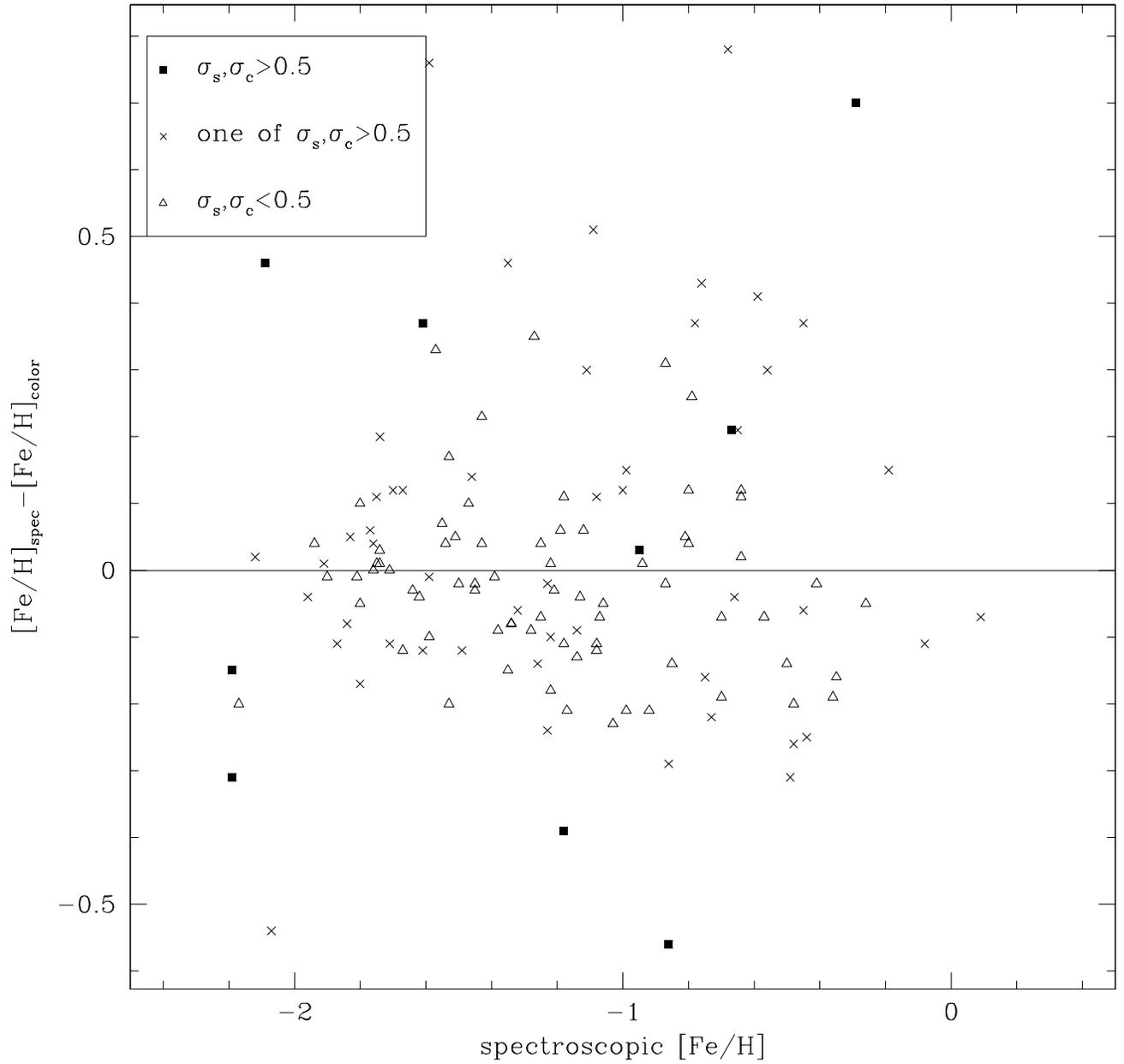}
\caption{Comparison of spectroscopic and color-derived metallicities for
M31 clusters with spectroscopic data. \label{metcol}}
\end{figure}

Eighty-seven of the cluster candidates in our sample of 221 with `reliable' reddening 
(as defined in the previous section)
have no spectroscopic information, so we attempted to estimate their
metallicities from their intrinsic colors.
We applied the BCES(Y$|$X) fits of metallicity
as a function of color, and averaged the resulting metallicities over all
available colors. As in the reddening determination, the standard deviation of
the metallicities from individual colors was used as the error estimate.
We tested this procedure by using it on the clusters {\it with} spectroscopic
information; this includes all the clusters used to do the metallicity-color
fits as well as additional objects with large metallicity errors.
The results are shown in Figure~\ref{metcol}; the mean offset
(spectroscopic$-$color-derived metallicity) is $0.020\pm0.021$, 
there is no evidence of a bias in the prediction with
metallicity, and the largest offsets are for objects with large
errors in color- or spectroscopically-determined metallicity or both. 

Applying the method to the clusters without spectroscopic data
produced equally encouraging results.
57\% of the color-derived metallicities had uncertainties 
${\sigma}_{\rm [Fe/H]} <0.5$, compared     
to 76\% within the same error range for spectroscopic metallicities.
Six objects had very large or small values of [Fe/H] ($>+0.5$ or $<-2.5$); 
these had only a few colors and large errors in their derived [Fe/H].
These objects do not lie on the same two-color sequences as the
confirmed globular clusters, so we suspect that they are either compact
background galaxies, compact H~II regions, or foreground stars.
We do not include these outlying metallicities in the analysis that follows.

\subsection{Color distributions}\label{sec-colordist}
\begin{figure}[t]
\includegraphics*[scale=0.9,angle=0]{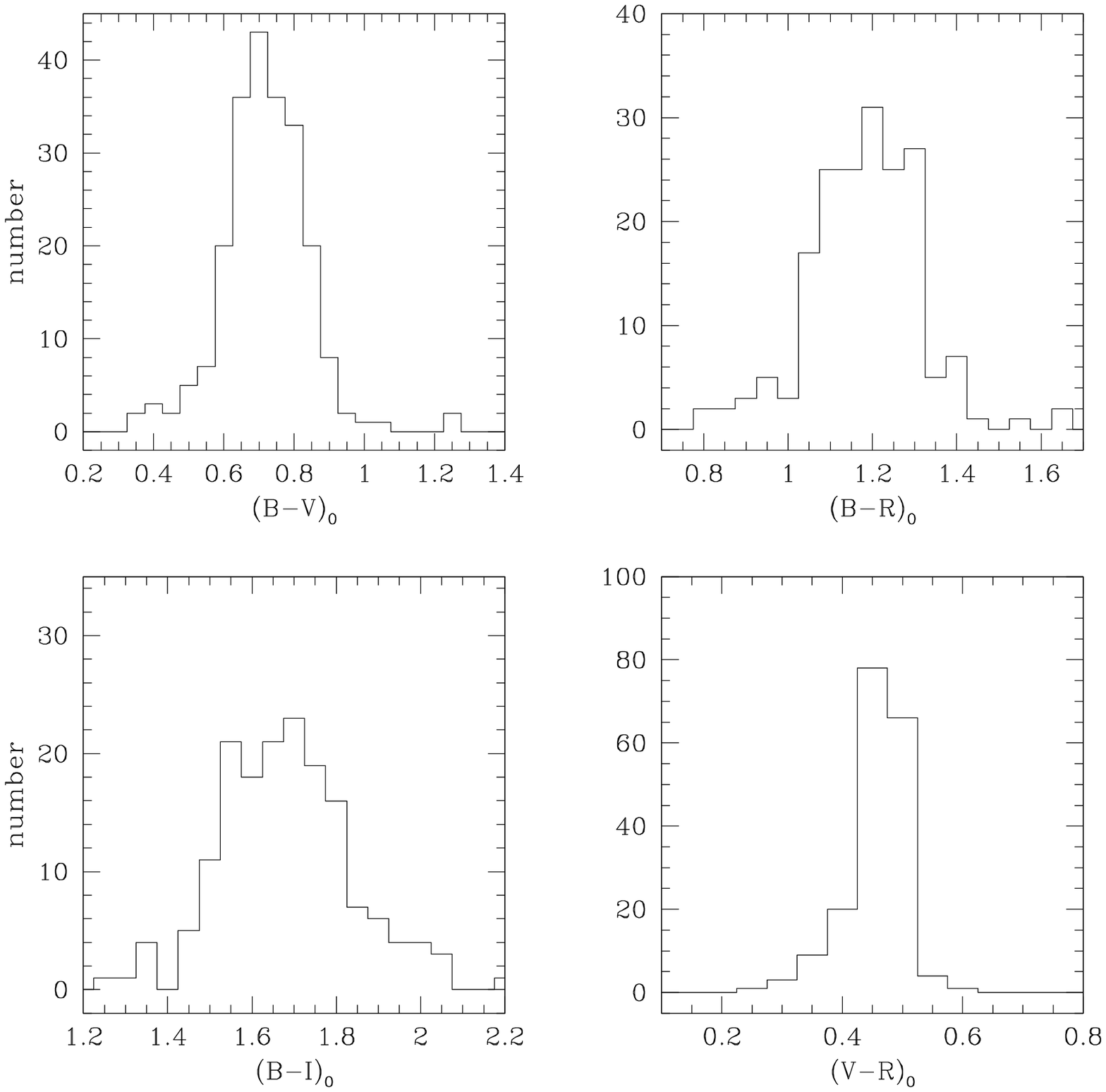}
\caption{Intrinsic optical color distributions for M31 GCs.\label{opt-hist1}}
\end{figure}

\begin{figure}[t]
\includegraphics*[scale=0.9,angle=0]{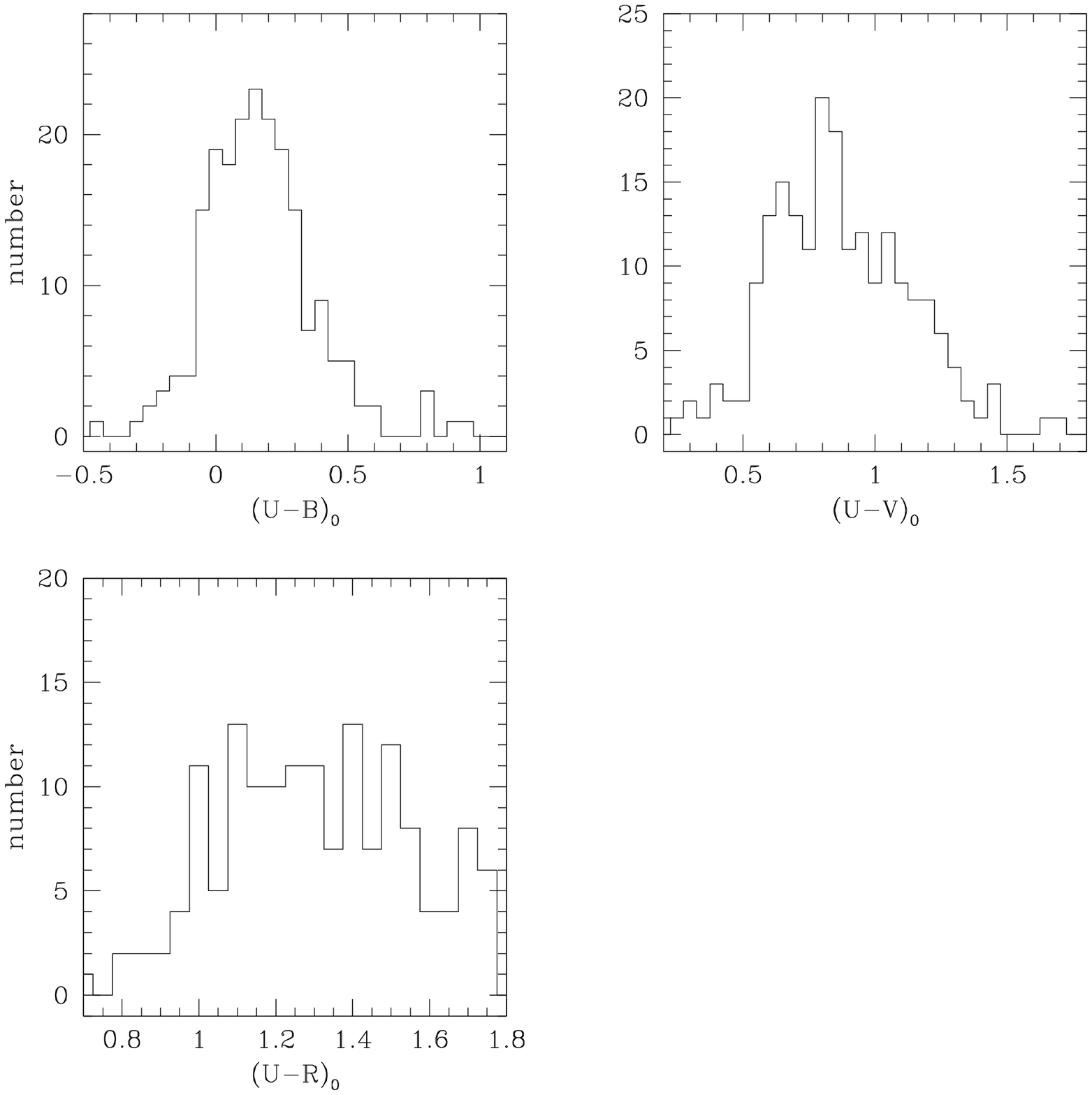}
\caption{Intrinsic optical color distributions for M31 GCs.\label{opt-hist2}}
\end{figure}

\begin{figure}[t]
\includegraphics*[scale=0.6,angle=0]{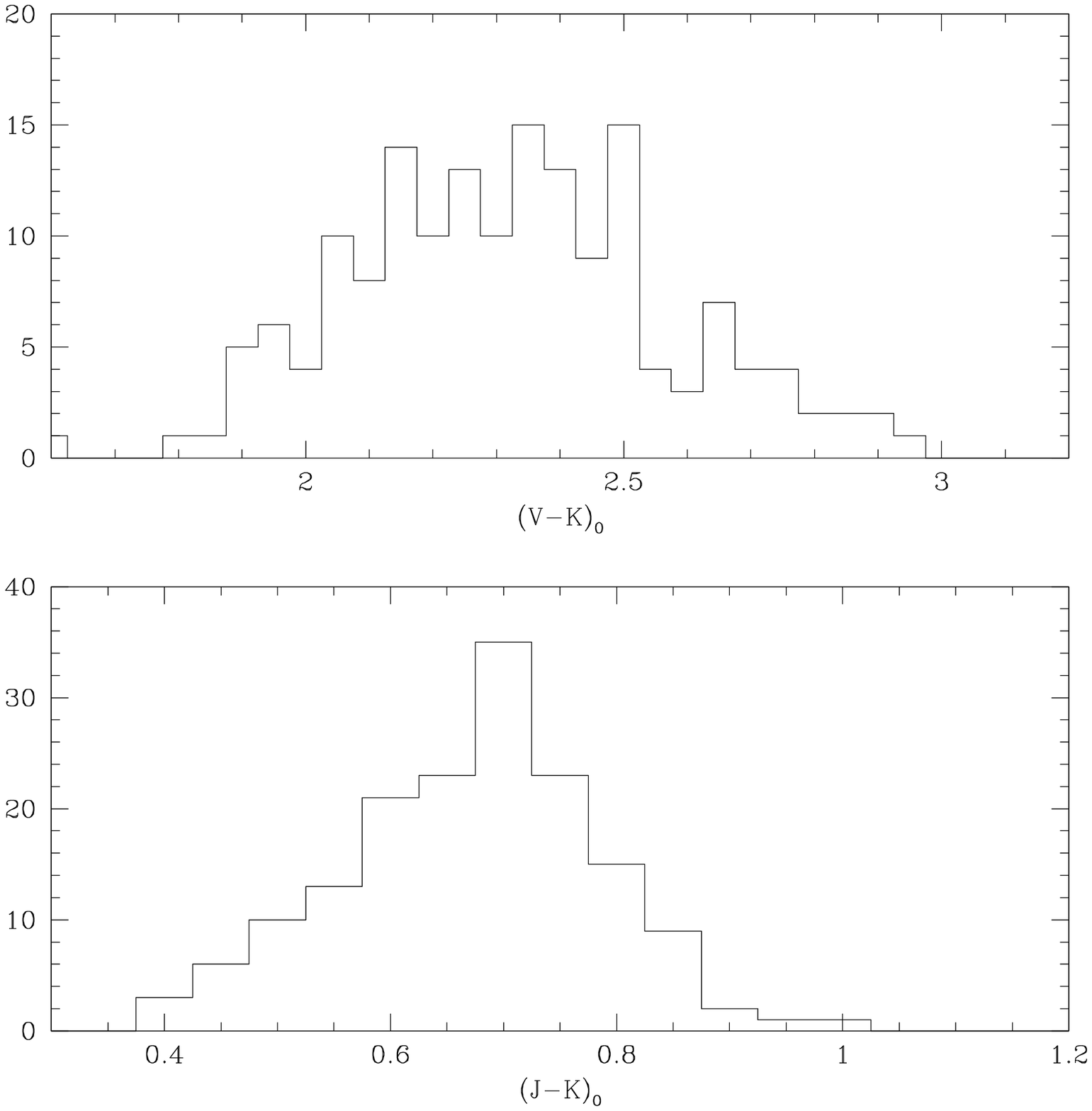}
\caption{Intrinsic optical-infrared color distributions for M31 GCs.\label{ir-hist}}
\end{figure}

We analyzed the distribution of intrinsic colors for the 221
M31 GCs with reliable reddening. The histograms of colors
are shown in Figures~\ref{opt-hist1}-\ref{ir-hist}, and parameters
of the color distributions are given in Table~\ref{tab-colordist}.
For comparison, the table also shows the mean intrinsic colors of the Galactic clusters
(optical from \citet{h96}, $(V\!-\!K)_0$ and $(J-K)_0$ from
\citealt{bh90}). The mean colors of the M31 clusters are 
consistent with corresponding Galactic mean colors.
The large standard deviations in the M31 cluster colors
incorporating $U$ probably reflect the larger photometric 
errors in this filter. 
$(V\!-\!K)_0$ is notable for having a larger range ($\sim 1.5^m$)
than most other colors. This is, of course, the basis 
for its use as a metallicity indicator.
$(V\!-\!R)_0$ is notable for having a very small range; as 
\citet{rhh92} reported, this can be exploited to
discriminate against background galaxies 
(which have $(V\!-\!R)_0 \gtrsim 0.7$) in cluster searches. 

We tested the color distributions of the M31 clusters for bimodality using the KMM
algorithm \citep{mb88,abz94}. The input to this algorithm includes the individual
data points, the number of Gaussian groups to be fit, and a starting point for
the groups' means and dispersions (the final solution is not very
sensitive to the starting points unless there are many outliers). 
We used the results of \citet{ab93} to choose our starting points:
they found two groups of M31 clusters with ${\rm [Fe/H]}=-1.5$ and $-0.6$, 
with the metal-poor clusters comprising two-thirds of the total.
Our input data specified two groups, with the bluer group twice
as large, and the two mean colors corresponding to ${\rm [Fe/H]}=-1.5,-0.6$
from our color-metallicity relations (Section~\ref{sec-colormet}).
The predicted mean colors for the two groups appear in the last two columns of
Table~\ref{tab-colordist}.
We specified the same dispersions for both groups in each color; 
in this case (`homoscedastic' fitting as opposed to `heteroscedastic') 
the $p$-value returned by KMM adequately measures the statistical significance of the 
improvement in the fit in going from one to two groups.
As a rough estimate, we specified a value of 80\% of the overall dispersion in 
Table~\ref{tab-colordist} as the starting point for the groups' dispersions
in each color.

The hypothesis of a unimodal color distribution was rejected for
only three colors: $(U\!-\!V)_0$ and $(U\!-\!R)_0$ at the $95$\% 
confidence level and $(V\!-\!K)_0$ at the $92$\% level.
The mean colors of the two groups in all three colors correspond 
to metallicities of approximately $-1.5$ and $-0.6$.
These three colors are the most sensitive to metallicity 
(Table~\ref{color-met}), so it would be expected that they 
would show the strongest evidence for bimodality.
In the other colors, the photometric errors are probably large enough
to mask any color separations between the metal-rich and metal-poor
populations.
Visual inspection of the color histograms suggested that these same three
colors might actually have trimodal distributions. We tested for
this, again using KMM, and found that three-group fits were not
superior to either one- or two-group fits for $(U\!-\!V)_0$ and $(U\!-\!R)_0$.
Three groups were preferred to one or two for $(V\!-\!K)_0$. We are reluctant to
claim a physical meaning for this, since this color is the most sensitive
to both photometric errors (separate optical and infrared photometry is
combined) and reddening. In the following section we show that
two metallicity groups are preferred.

\begin{figure}[t]
\includegraphics*[scale=0.9,angle=0]{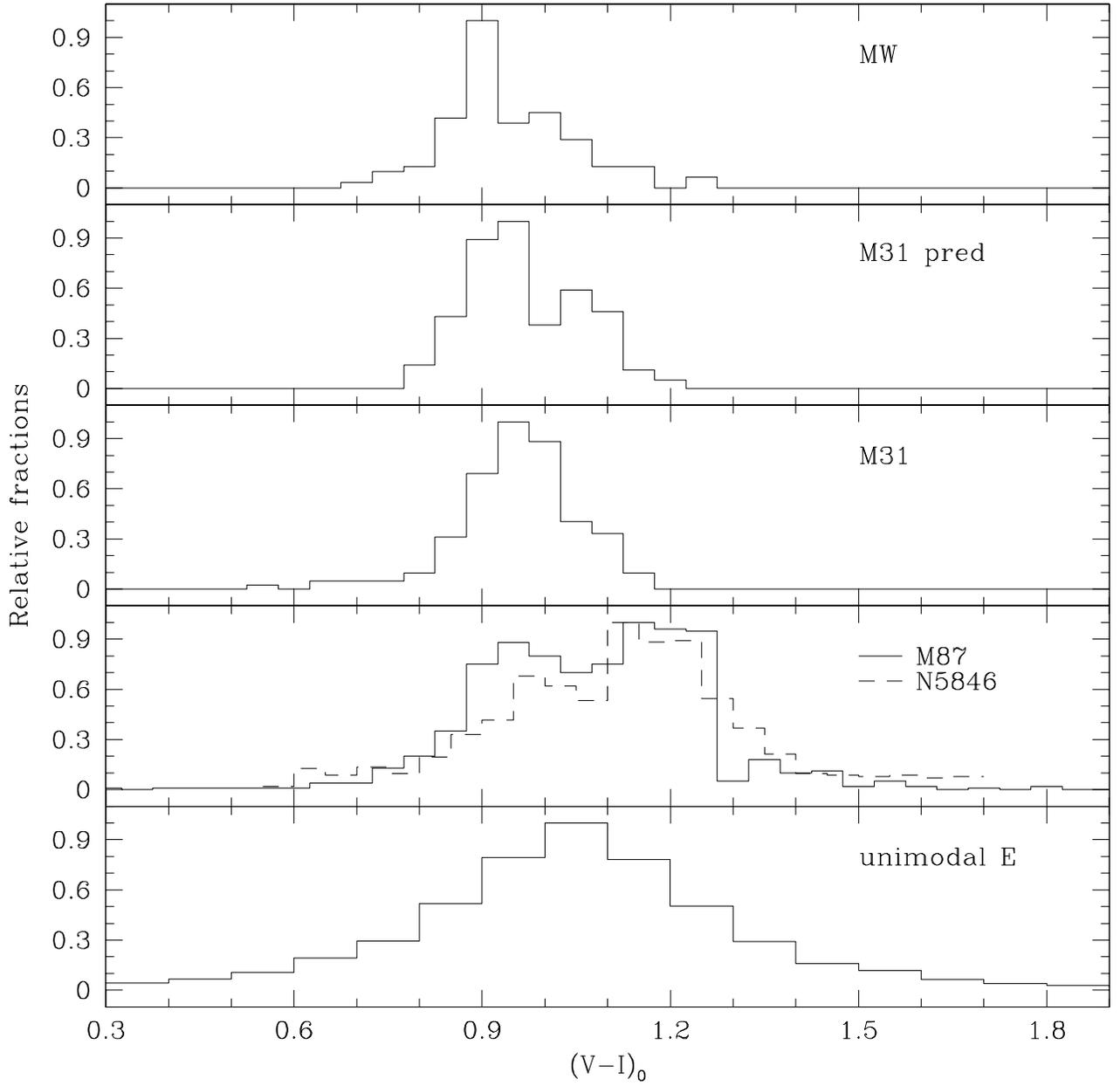}
\caption{Distribution of $(V-I)_0$ for GCs of several galaxies.`M31 pred' refers
to $(V-I)_0$ predicted from [Fe/H] of M31 GCs.\label{vi-hist}}
\end{figure}

Figure~\ref{vi-hist} shows the distribution of $(V\!-\!I)_0$,
which is often used as a metallicity
indicator for globular cluster systems despite its
fairly low metallicity sensitivity (see Table~\ref{color-met}). 
From HST imaging in $V$ and $I$, \citet{kun99} finds that
25-50\% of the GCSs of a sample of $\sim50$ galaxies show
evidence for bimodal color distributions; \citet{gkp99} find
similar results.
The bottom two panels of the figure show the color distributions
for elliptical galaxy GCSs with and without bimodality.
The M87 \citep[data from][]{kws99} and NGC~5846 GCs \citep[data from][]{for97}
clearly show bimodal distributions in $(V\!-\!I)_0$.
The `unimodal E' panel is the sum of 12 elliptical GCS color distributions
which \citet{for96} find {\em not} to be bimodal; although 
the histogram bins are larger in these data,
the distribution is remarkably symmetric and unimodal.
Comparing the color distribution for the ellipticals and
spirals yields two interesting conclusions: first, M31
and the Galaxy clearly lack the extremely red (and
presumably metal-rich) GCs found in massive ellipticals.
Second, the blue peaks of the M87 and NGC~5846 color distributions
are at approximately the same color as the M31 and Galactic peaks.
This is consistent with the finding that
these galaxies' metal-poor GC populations and the total M31
GC population have approximately the same mean metallicity
(${\rm[Fe/H]}\approx -1.2$; \citet{fbg97} and the following section),
and is also an interesting hint of a possible connection between 
ellipticals' metal-poor GCs and spirals' GCs.

The Galactic GCs \citep[see, e.g.][]{cot99}
and the M31 GCs \citep[and the following section]{ab93}
are known to have bimodal metallicity distributions, and we 
have just shown that some M31 cluster colors are bimodal --
why not $(V\!-\!I)_0$? We did Monte Carlo simulations of our 
observations of the \vi\ distribution, and, as suggested 
above, we found that observational errors in the
reddening and photometry can wash out the signature
of bimodality. We predicted the M31 GCs' `true' $(V\!-\!I)_0$
colors from their spectroscopically-determined [Fe/H] (see the middle
panel in Figure~\ref{vi-hist}); we found this `true' distribution 
to be bimodal at the $>99$\% confidence
level. We then added to each color datum a Gaussian random 
error, drawn from a distribution with mean of 0 and standard
deviation expected from the errors in our photometry and reddening 
determination.
Of the 1000 color distributions generated in this manner, KMM
detected bimodality in only about 250, implying that 
observational errors wash out the bimodal signal three-quarters
of the time. Detection of multiple populations in GCS color
distributions thus clearly requires precise photometry and/or
the use of metal-sensitive colors.

We examined the correlation of M31 cluster 
intrinsic colors with distance from the galaxy's center, 
using the coordinate system of \citet{ba64}, as defined in HBK.
In this system, $X$ is the projected distance from the center of M31 along the
major axis (positive $X$ is to the northeast), $Y$ is the projected distance 
along the minor axis (positive $Y$ is to the northwest), and $R$ is the 
projected radial distance from the galaxy center, $R=\sqrt{X^2+Y^2}$.
Ideally we would use the true spatial distance and not the projected
distance from M31, but this information is not available for the M31 GCs.
We binned the clusters in 20\arcmin\ bins in $X$ and $Y$ and 10\arcmin\
bins in $R$, then calculated the weighted least-squares fit of the 
bin median colors against distance. It is well-known \citep{ir85,ew88}
that the observed colors of M31 clusters are redder for $Y>0$, 
because the northwest side of the M31 disk is closer to us and more
clusters are projected behind it. If our reddening correction 
was adequate this trend should be removed from the intrinsic colors.
None of the colors showed a significant trend with $X$ or $Y$,       
confirming that our reddening correction worked. More surprising was
the fact that none of the colors showed a significant trend with $R$;
such a trend would be expected if there was any gradient in the
metallicity of the system. However, even a large metallicity
gradient (for example, 0.5 dex over 100\arcmin) would produce a
fairly small change in most colors ($\lesssim 0.15^m$) so perhaps
photometric errors and scatter within the bins mask any true gradient.

Comparing the clusters' location in two-color diagrams to models produced by
population synthesis provides useful checks on our photometry and
on the models' accuracy. We obtained predicted
colors for populations of ages 8 and 16 Gyr from three sets of models:
\citet{w96}, \citet{bc96} (hereafter BC), and \citet{kff99} (hereafter KFF).
We used \citeauthor{w96}'s `vanilla' models and his interpolation program to
generate colors for values of [Fe/H] from $-2.0$ to $-0.1$ in steps of 0.1 dex.
We used the Salpeter IMF versions of the \citeauthor{bc96} and \citeauthor{kff99} 
models, without interpolation, and obtained colors at metallicities of 
$-2.33$ (KFF models only), $-1.63$, $-0.63$, $-0.32$, 0.07 and 0.47 dex. 
\citet{wor94} states that, compared to Galactic GCs, his models are too red 
by 0.08$^m$ in \bv\ and too blue by 0.03$^m$ in \jk, so we corrected the model
colors by these amounts. Worthey attributes these offsets to defects in the 
stellar flux library; since all three sets of the models share the same stellar 
atmosphere models we applied the same corrections to the BC and KFF models.

\begin{figure}[h]
\includegraphics*[scale=0.9,angle=0]{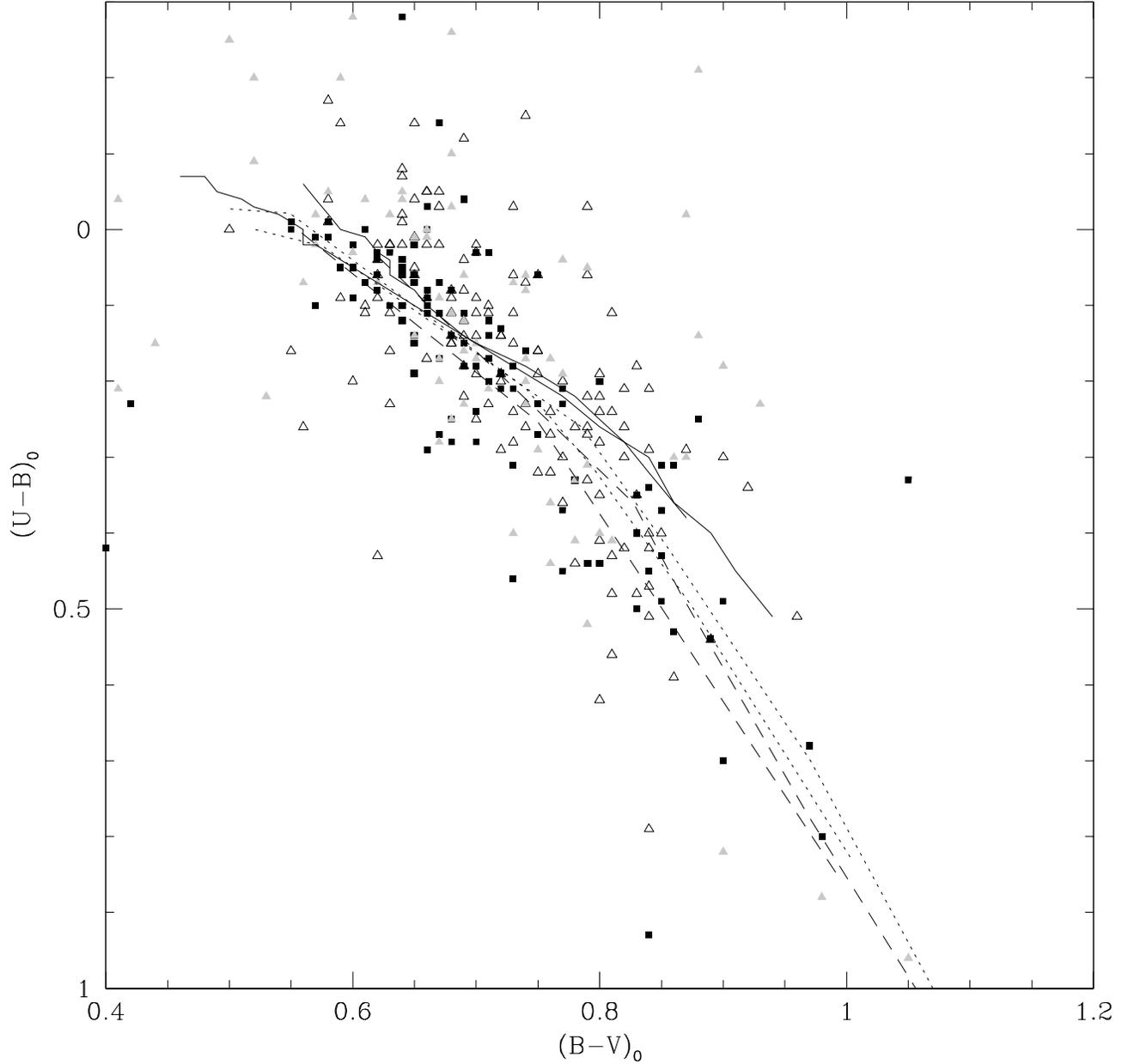}
\caption{$(B-V)_0$ vs. $(U-B)_0$ for Galactic GCs (squares), 
confirmed M31 GCs (open triangles), and M31 GC candidates (shaded triangles).
Lines are population synthesis models of ages 8~Gyr (bluer colors) and 16~Gyr (redder colors): 
\protect{\citet{w96}} (solid), \protect{\citet{bc96}} (dashed), 
\protect{\citet{kff99}} (dotted). Models have been corrected as described in the text.\label{bv_ub}}
\end{figure}

\begin{figure}[h]
\includegraphics*[scale=0.9,angle=0]{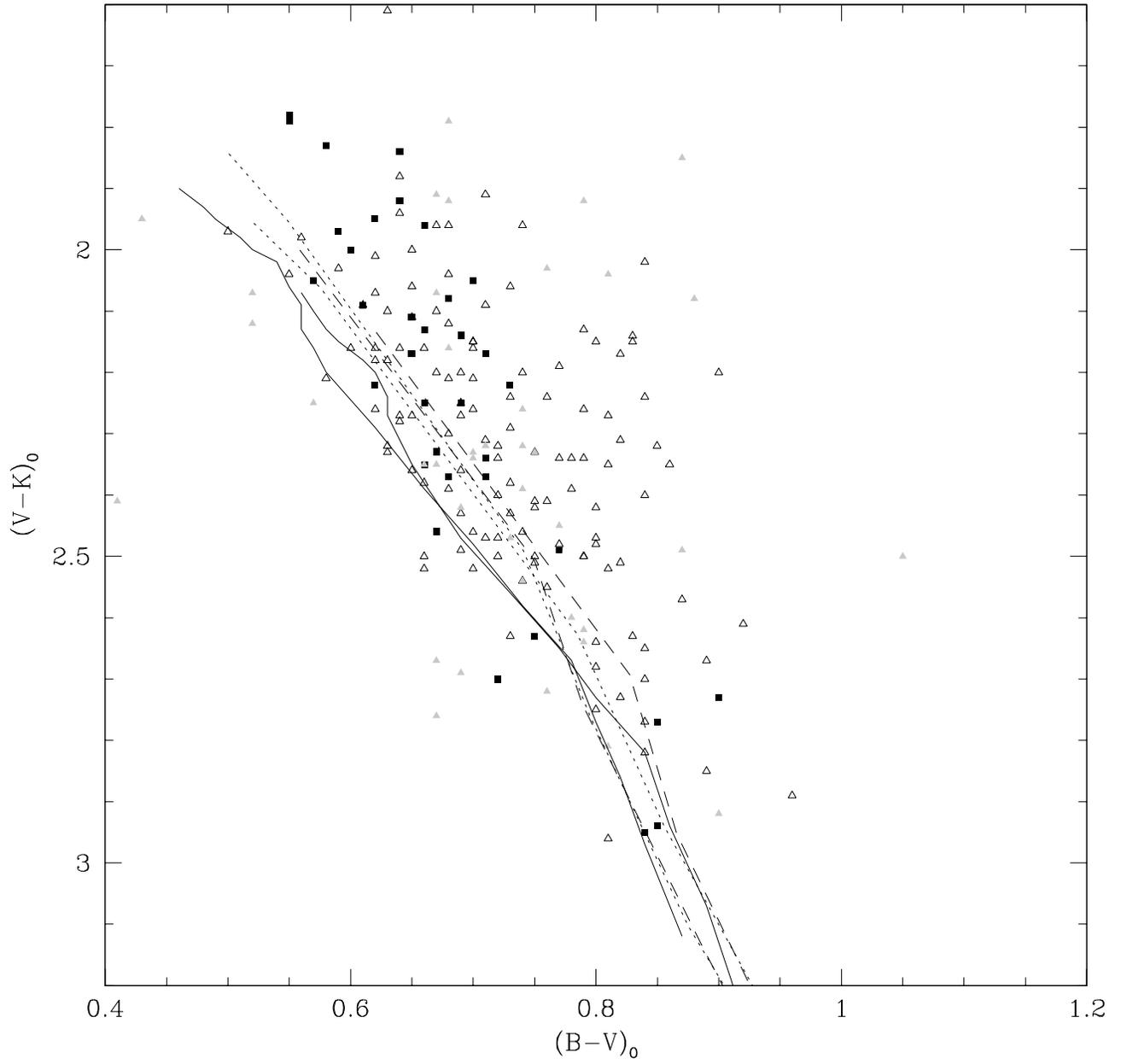}
\caption{$(B-V)_0$ vs. $(V-K)_0$ for M31 and Galactic GCs. Symbols as in
Figure~\ref{bv_ub}.\label{bv_vk}}
\end{figure}

\begin{figure}[h]
\includegraphics*[scale=0.9,angle=0]{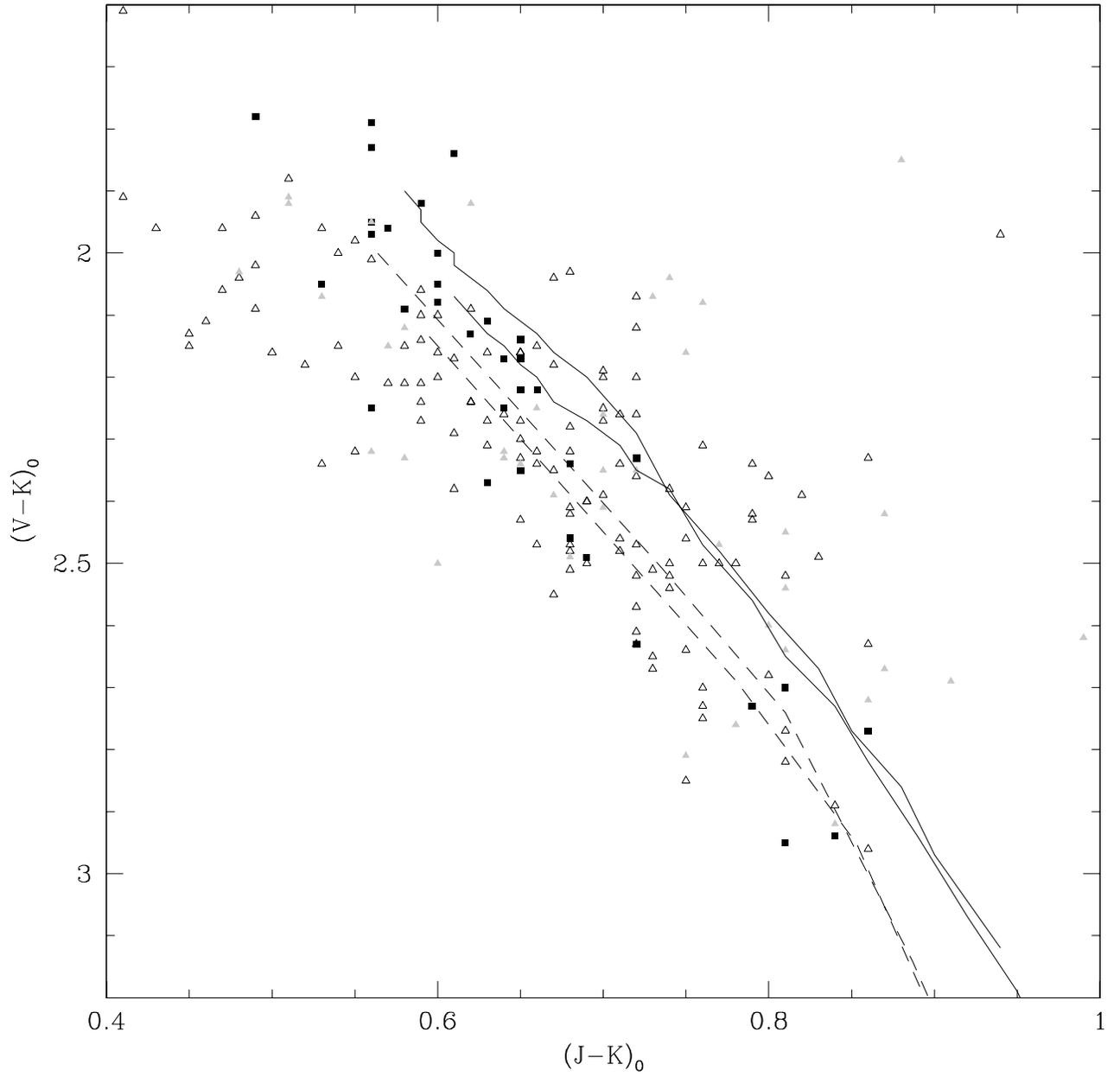}
\caption{$(J-K)_0$ vs. $(V-K)_0$ for M31 and Galactic GCs. Symbols as in
Figure~\ref{bv_ub}; the KFF models do not predict $J$ so are absent from this figure.\label{jk_vk}}
\end{figure}

In Figures~\ref{bv_ub}-\ref{jk_vk} we plot two-color diagrams for 
M31 clusters, Galactic clusters and the models, using optical 
and IR colors often found in the literature. 
Confirmation that our photometry suffers no major systematic 
errors is provided by the fact that the Galactic and M31 clusters
lie on essentially the same loci in all the diagrams.
As expected, the M31 clusters show much more scatter than the
Galactic GCs (because the photometric and reddening errors are larger),
but much of this scatter is due to objects that are not confirmed clusters.
It is clear from the diagrams that integrated photometry and model
predictions are not precise enough to distinguish any possible age differences between the 
two sets of old clusters.

In Figure~\ref{bv_ub} the corrected models agree reasonably well
with the data in \bv\ and \ub, although the agreement becomes poorer
in the high-metallicity region.
The models also agree fairly well with each other in these colors,
which is not the case in Figures~\ref{bv_vk} and \ref{jk_vk}.
The model disagreement is not surprising; \citet{cwb96} found a difference of $0.3^m$ in
predicted \vk\ between the solar-metallicity models of \citeauthor{w96}
and those of \citeauthor{bc96}. These authors attribute most of the
discrepancy to differences in the underlying stellar evolution
prescriptions. 
Shifting the Worthey models by $\sim 0.2^m$ to bluer \vk\ 
to match the data in Figure~\ref{bv_vk}
(since Figure~\ref{bv_ub} implies that the \bv\ color is acceptable) would then require
shifting the same models to bluer \jk\ by $\sim 0.1^m$ to match the data in
Figure~\ref{jk_vk}. 
The shifts required for the BC and KFF models to fit the data
in the \bv/\vk\ and \jk/\vk\ planes are in the same direction, but
about half the magnitude, as those required for the Worthey models.
This arbitrary shifting of predicted colors to match the 
data does not uniquely determine the reasons for the disagreements
between models and data. We speculate that the mismatches may be
caused by problems in the treatment of the late stages of stellar evolution 
in the models, since most of the $K$-band light comes from evolved stars.
The theoretical and observation photometric systems could also have systematic
offsets. Clearly this is an area requiring more detailed examination by modelers,
and the colors of globular clusters provide important
constraints on the models.

\begin{figure}[h]
\includegraphics*[scale=0.9,angle=0]{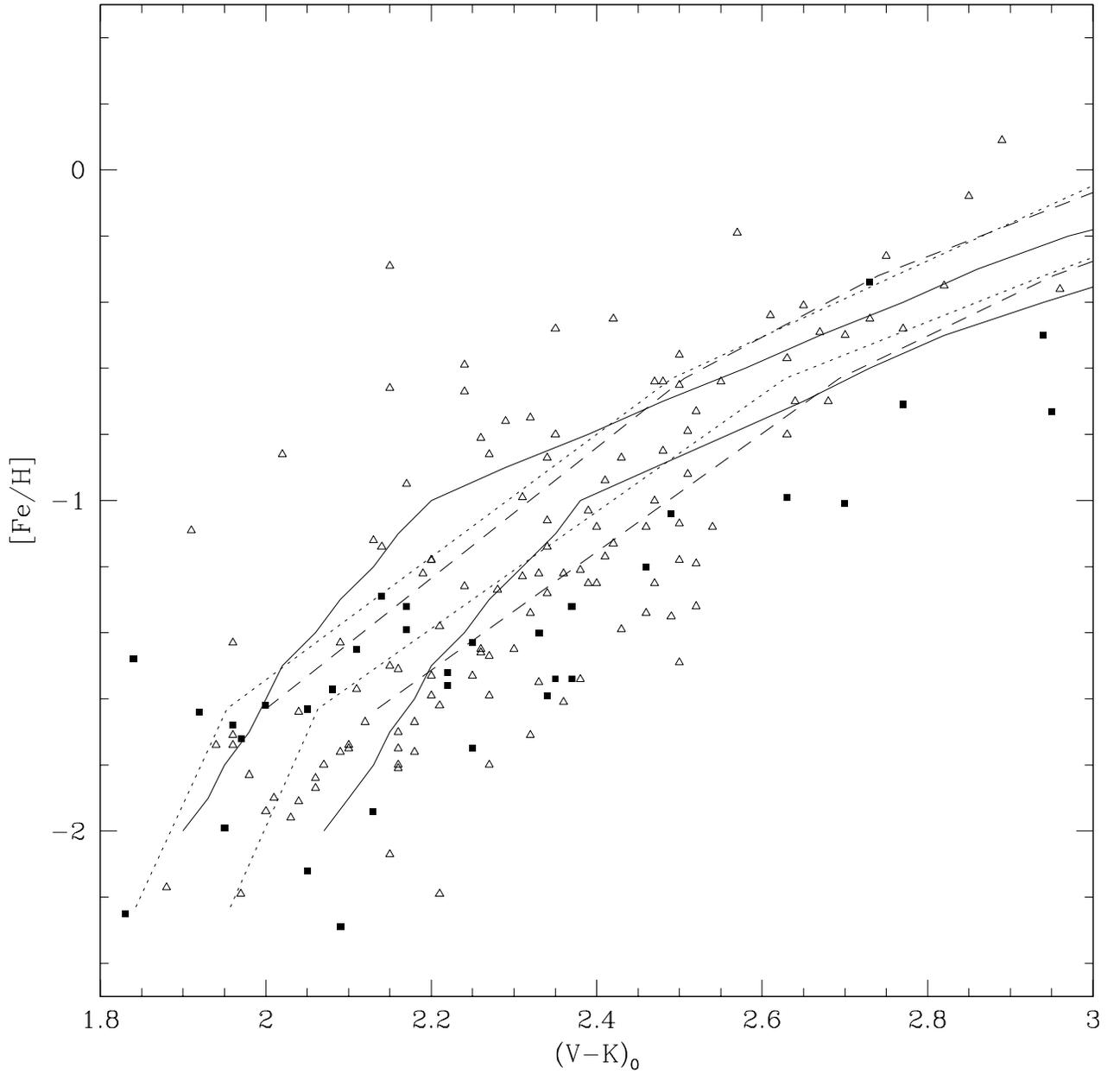}
\caption{$(V-K)_0$ vs. [Fe/H] for M31 and Galactic GCs. Symbols same as
Figure~\ref{bv_ub}.\label{vk_feh}}
\end{figure}

In Figure~\ref{vk_feh} we plot [Fe/H] as a function of $(V\!-\!K)_0$.
Some M31 GCs are too blue for their metallicities or too metal-rich
for their colors, as discussed in Section~\ref{sec-colormet}.
\citet{cov94} noted that some low-metallicity Galactic clusters 
were ``exceedingly blue'' with respect to the models of \citet{buz89}; 
they attribute this to a systematic problem with the photometry.
\footnote{\citeauthor{cov94} incorrectly attribute the
Galactic cluster photometry to \citet{bh90}. While the
photometry is tabulated in the \citeauthor{bh90} paper, 
the actual data are from \citet{fpc80}.} 
However, we find that the M31 clusters and Galactic clusters overlap 
in the region $(V\!-\!K)_0\lesssim2.1$ (Figure~\ref{vk_feh}),
and that the $(V\!-\!K)_0$-[Fe/H] 
relations for the two sets of clusters are very similar at
the metal-poor end (Figure~\ref{ir_colormet}). 
This implies that the \citet{buz89} models are too red, 
rather than the clusters being too blue, and indeed the \citeauthor{buz89}
models are about $0.1^m$ redder than the models we examine here.
We showed in the previous paragraph that substantial shifts
in model \vk\ colors were required to match the clusters in two-color
diagrams. Shifting the model \vk\ to the blue to match the \bv\ colors
would make the model \vk\ too blue for the clusters' metallicities.
This again emphasizes the difficulties of trying to match 
population synthesis models to cluster colors by applying
uniform shifts for all metallicities.

\subsection{Metallicity distributions}\label{sec-metdist}
The metallicity distribution of a galaxy's GCS can provide
important clues to galaxy formation. For example, 
\citet{zin93} finds a significant metallicity gradient in the
Galactic `old halo' clusters and no gradient in the `younger halo'.
He interprets this and other properties of the Galactic GCS as evidence that the
old clusters were formed in a monolithic collapse and the 
younger ones were accreted from satellite galaxies.
Accordingly, we want to examine the distribution of cluster 
metallicities in M31, and to do so for the largest number of clusters. 
We thus include metallicities estimated from colors in our analysis, even though 
this method is not as precise as determining metallicity 
spectroscopically. We also include clusters for which we were unable to
determine a reliable reddening value (usually because of inadequate photometric
data) in the metallicity analysis.

\begin{figure}[t]
\includegraphics*[scale=0.9,angle=0]{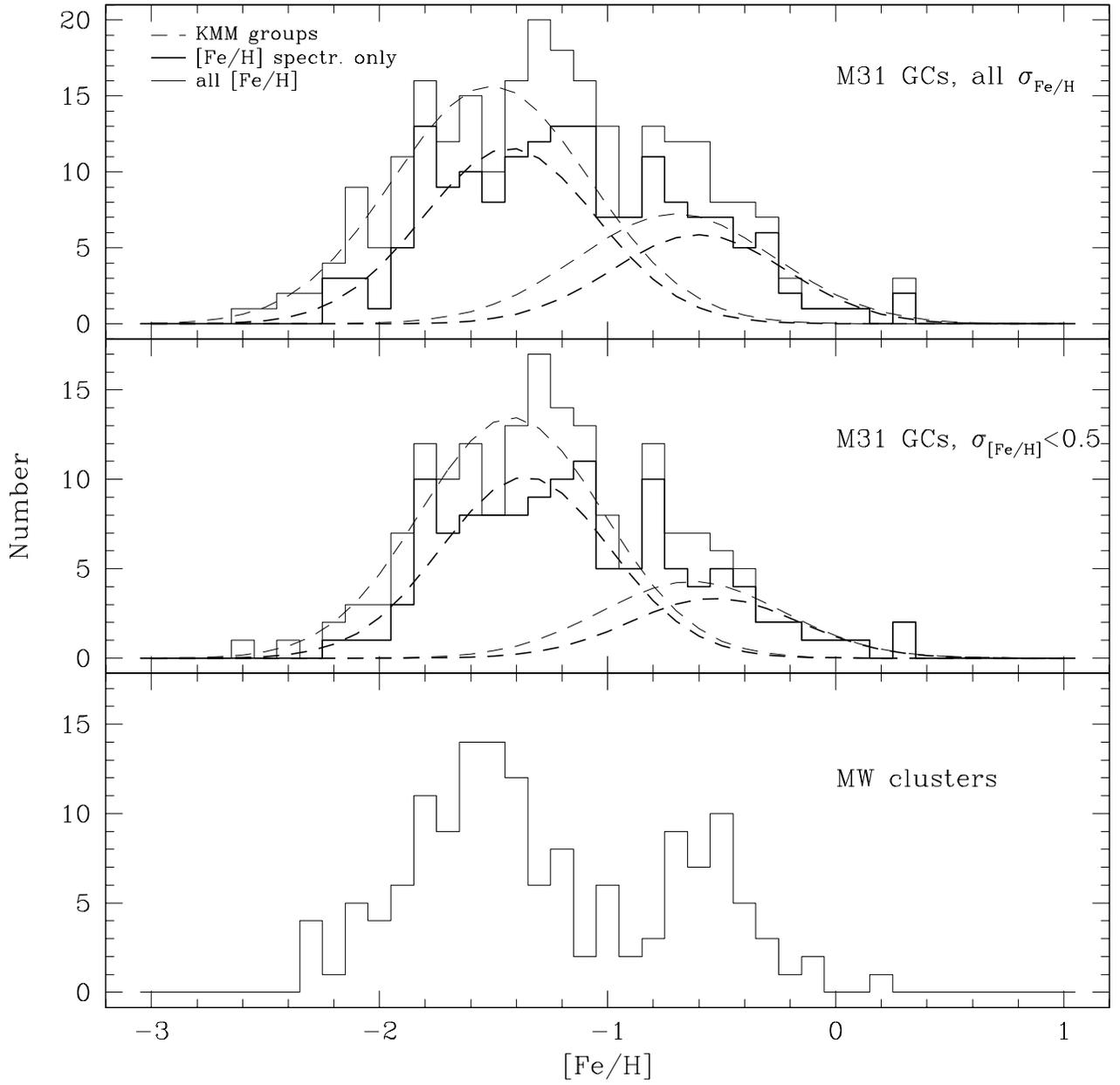}
\caption{[Fe/H] distribution for M31 GCs, subdivided by metallicity
source and uncertainty, and Galactic GCs.\label{fehhist}}
\end{figure}

To assess whether the determination of metallicities from colors
has an effect on our results, we consider four data sets in
our analysis of the metallicity distribution.
Set 1 contains all the objects for which metallicities
have been determined, regardless of method or error. Set 1a is a subset
of this, containing only objects with ${\sigma}_{\rm [Fe/H]} <0.5$.
Set 2 contains only objects with spectroscopic metallicities, and
set 2a is the subset of these objects with ${\sigma}_{\rm [Fe/H]} <0.5$.
The first step is to characterize the distribution 
of [Fe/H]. Table~\ref{fehdist} shows that restricting the metallicities
to spectroscopic alone slightly increases the mean metallicity, but  
not by a significant amount. This is good evidence that our 
color-derived metallicities are not systematically offset from the
spectroscopic ones. A KS-test shows that the Galactic and M31 GC     
metallicity distributions are not drawn from the same distribution,
unsurprising given the difference in mean metallicity; however, the shapes
of the distributions are fairly similar (see Figure~\ref{fehhist}).

The asymmetric nature of the [Fe/H] distributions suggests the
possibility of bimodality. We used the KMM algorithm to search for
bimodality in the distributions, following Ashman and Bird as in 
Section~\ref{sec-colordist}. All four datasets showed bimodality,
although including the color-derived metallicities made the
detections only marginally significant. (see the $p$-values in Table~\ref{fehdist}).
The [Fe/H] histograms and the Gaussian subgroups found by KMM 
are shown in Figure~\ref{fehhist}.
Most clusters were assigned to the same (metal-rich or metal-poor) group
regardless of which dataset was considered; however, there were
about 20 clusters which showed substantial probability of membership in
both groups and these switched groups from metal-rich to metal-poor
depending on whether or not the large-error metallicities were included. 
This is the primary cause of the differences between 
samples 1/1a and 2/2a in Table~\ref{fehdist}.
We conclude that these clusters cannot be unambiguously assigned
on the basis of their metallicities alone, so we classify them as
`intermediate' and assign them to neither group.
As with the color distributions, visual inspection of the 
metallicity histograms suggested the possibility of trimodality
in the distribution. The results of running KMM with three groups
specified were similar to those for the color distributions:
three-group fits were worse than both one- and two-group fits.
The mean colors of the metal-rich and metal-poor groups are
generally within $0.05^m$ of the predicted colors in Table~\ref{tab-colordist}. 
This is not surprising since the metallicities used to predict these
colors are close to the mean metallicities of the two groups KMM found.
Perhaps more surprising is our failure to detect 
bimodality in most colors. This underscores the need for precise
photometry if a single color is used to determine metallicity:
such precision was obviously not achieved in our heterogeneous
data set.

\begin{figure}[t]
\includegraphics*[scale=0.9,angle=0]{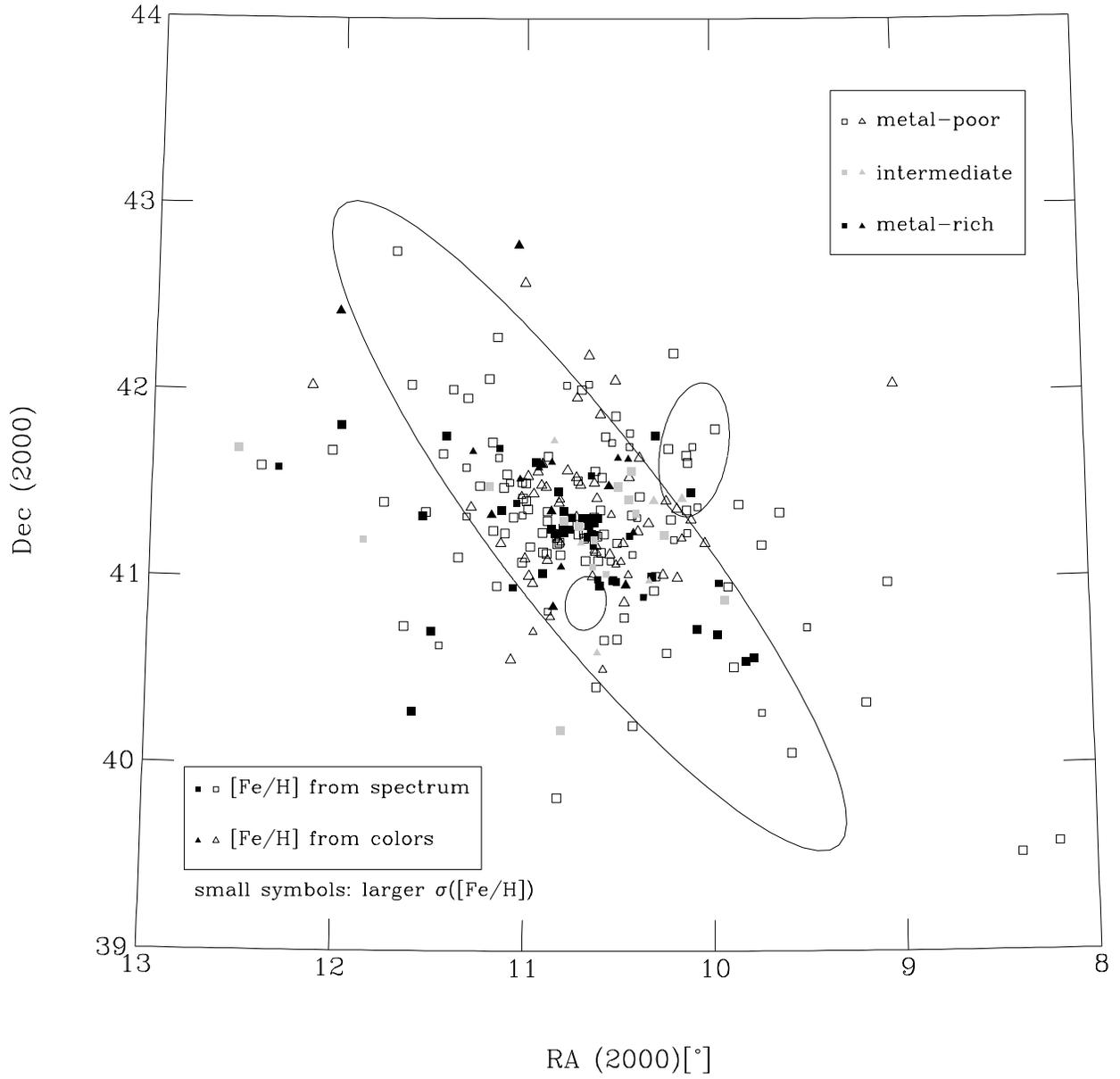}
\caption{Projected location of metal-poor and metal-rich M31 clusters. 
Ellipses same as Figure~\ref{redmap}.
\label{fehgroups}}
\end{figure}

Do these metal-rich and metal-poor groups represent the
M31 equivalent of the Galactic disk (or bulge -- see \citealt{cot99}) 
and halo clusters?
One way to find out is to see where the two groups
lie in relation to the galaxy; we show the KMM assignments
function of position on the sky in Figure~\ref{fehgroups}. 
There are more metal-poor clusters at large projected
radius, and as a result the median radius of the metal-poor
clusters is about 50\% larger than that of the metal-rich clusters. 
However, we find, as did HBK,
that this is partly a selection effect: faint clusters are
more easily discovered away from the disk, and these 
distant clusters are more likely to be metal-poor. 
When the sample is magnitude-limited at $V=16.5$ or $V=17$ 
we find that the median radius of
the metal-poor clusters is about 30\% larger than that of the
metal-rich clusters, which is similar to the results of HBK. 
While most of the metal-rich clusters are projected onto the
M31 disk, a few lie in the halo. Two of these clusters have color-magnitude 
diagrams (379-312 from \citet{cmd3} and 006-058 from \citealt{cmd5}) that
give values of [Fe/H] consistent with our spectroscopic 
values. The CMD of 384-319 \citep{cmd2} is so sparse 
that it cannot put any useful constraints on [Fe/H]. The existence
of several metal-rich globular clusters in the Galactic halo
(Terzan~7, NGC~6366, Pal~12; \citealt{dca95}) also suggests that
similar results for the M31 clusters are not unreasonable.

The kinematics of the metal-rich and metal-poor groups 
will be the most powerful determinant of their similarity (or lack thereof)
to the two Galactic groups.
We did not perform a detailed kinematical analysis, since  
two forthcoming velocity studies of the M31 clusters \citep{sei99,per99} 
have substantially improved precision from HBK's observations 
(the source of over three-quarters of our velocity data).
We did repeat the HBK analysis (see their Figure~6 and Table~3), 
including our new velocities and using our division of the clusters, 
and found essentially the same results. At small projected distances,
the metal-rich clusters rotate faster than the metal-poor clusters,
and at large distances, there is essentially no difference between the two groups
in either rotation velocity or velocity dispersion.
The rotation velocity for all clusters at $X>10$\arcmin\ is $59\pm12$ km~s$^{-1}$.
The similarity in the kinematics of the metal-rich and
metal-poor clusters hints that, unlike the Galactic clusters,
the two groups of M31 clusters might be similar in age.
Velocity errors and projection effects could confuse the
situation, however, and more precise velocities and metallicities 
are needed.

How does the distribution of M31 GCS metallicity compare to 
that seen in other galaxies? Table~\ref{gcscomp} compares 
some properties of spiral galaxies' globular cluster systems. 
Spirals with detected GCSs which do not appear in this
table (see Appendix of \citet{az98} and also \citealt{har91}) 
have generally been observed in only one filter, so no metallicity
information is available for these systems.
M31 and the Galaxy are the only spiral GCSs for which the 
detection of multiple populations is reasonably secure, and the populations
in these two galaxies are quite similar in metallicity difference
and relative proportion.
The small number of GCs in the other spirals' GCSs makes analysis of the
metallicity distribution difficult. For the M81 GCs, Figure~16 of \citet{pr95}
hints at a multimodal (or perhaps uniform plus one peak) distribution
of $(B-R)_0$. However, many of the objects in this plot were 
subsequently determined to be non-clusters \citep{pbh95}, and the 
number of remaining {\it bona fide} clusters is again too small for the
distribution to be analyzed. \citet{bri97} do not determine
individual metallicities from their spectra of GCs in NGC~4594 (M104), but
\citet{fgs97} do find some evidence for a difference in mean
$B-I$ color between disk and bulge/halo clusters in this galaxy.

\begin{figure}[t]
\includegraphics*[scale=0.9,angle=0]{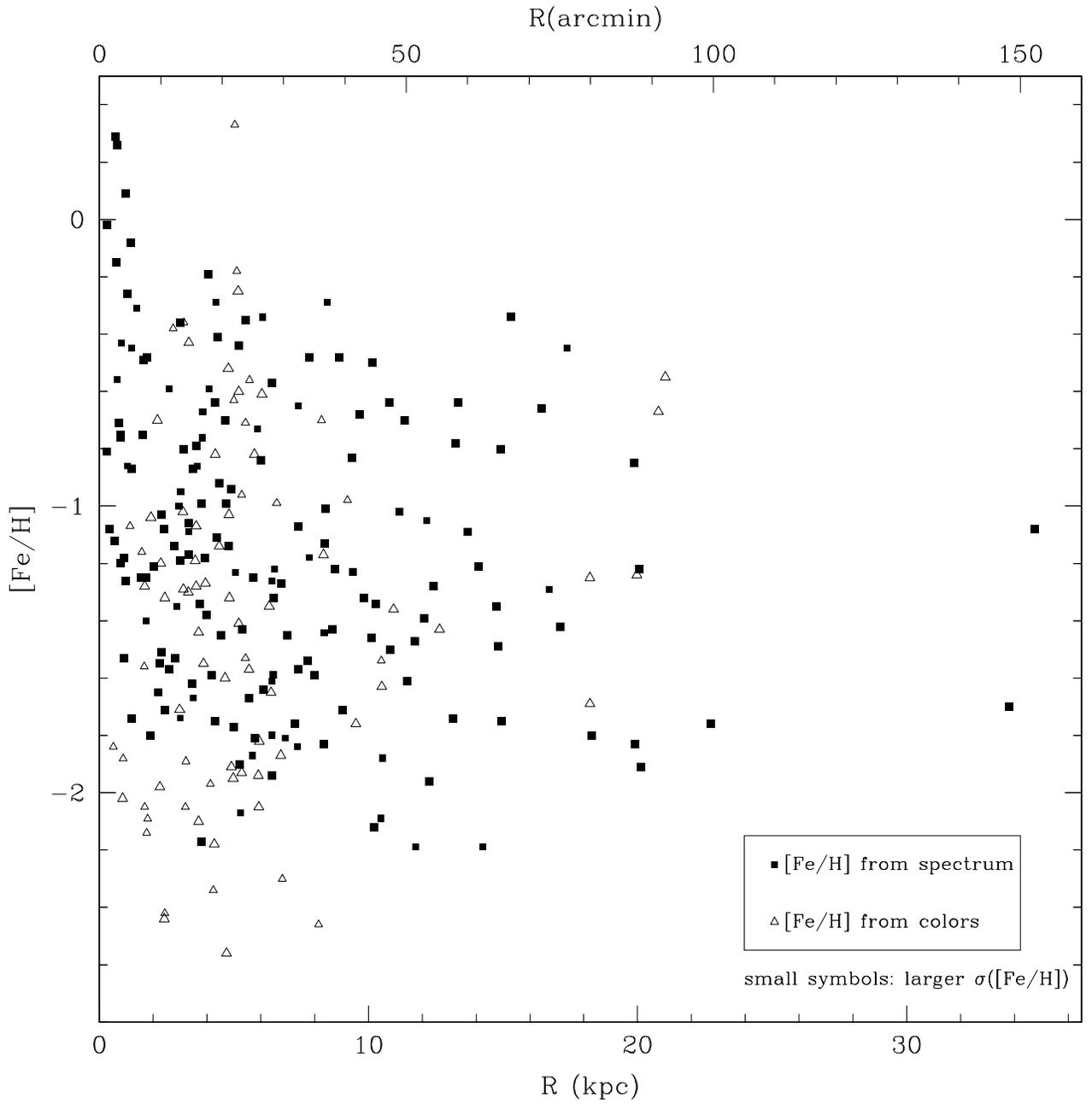}
\caption{[Fe/H] vs. $R$ for M31 clusters.\label{feh_r}}
\end{figure}

The presence or absence of a radial trend in GCS metallicity is
an important test of galaxy formation theory. 
We show the GC metallicity as a function of projected 
radius in Figure~\ref{feh_r}.\footnote{The 
absence of clusters with 100\arcmin$<R<150$\arcmin\ 
in Figure~\ref{feh_r} is a selection effect. The \citet{bat87} 
catalog extends to $R\sim120$\arcmin, and the \citet{sar77} 
extends to $R\sim155$\arcmin\ only along the M31 minor axis, so most of the 
region 120\arcmin$<R<150$\arcmin\ has not been searched for GCs.}  
This figure shows that the most metal-rich clusters
are near the M31 nucleus, and also shows the decrease in the `upper 
envelope' of GC metallicity noted by HBK.
We binned the clusters in distance as in Section~\ref{sec-colordist}, and 
looked for trends of metallicity with radial distance $R$. 
The entire sample of clusters does not have a significant 
radial metallicity gradient, but the clusters with 
spectroscopic metallicities have a marginally
significant gradient ($-0.023\pm0.01$ dex/kpc). 
This metallicity gradient is close to the value \citet{zkh94}
find for the [O/H] gradient in M31 HII regions ($-0.018\pm0.006$ dex/kpc),
emphasizing that the properties of M31 and its GCS are closely
linked. The metallicity gradient of the Galactic halo clusters 
\citep[excluding six distant clusters ][]{adz92}, $-0.011\pm0.004$ dex/kpc,
is smaller than the M31 GCS gradient, probably because
the M31 sample includes the very metal-rich clusters in the nucleus. 
We cannot determine the metallicity gradients for M31 `disk'
and `halo' clusters separately because using metallicity itself to 
assign the M31 clusters to disk/halo groups would strongly
bias the result. Better kinematical data will allow an
independent group assignment, and hence better understanding
of the groups' properties. 

\begin{figure}[t]
\includegraphics*[scale=0.9,angle=0]{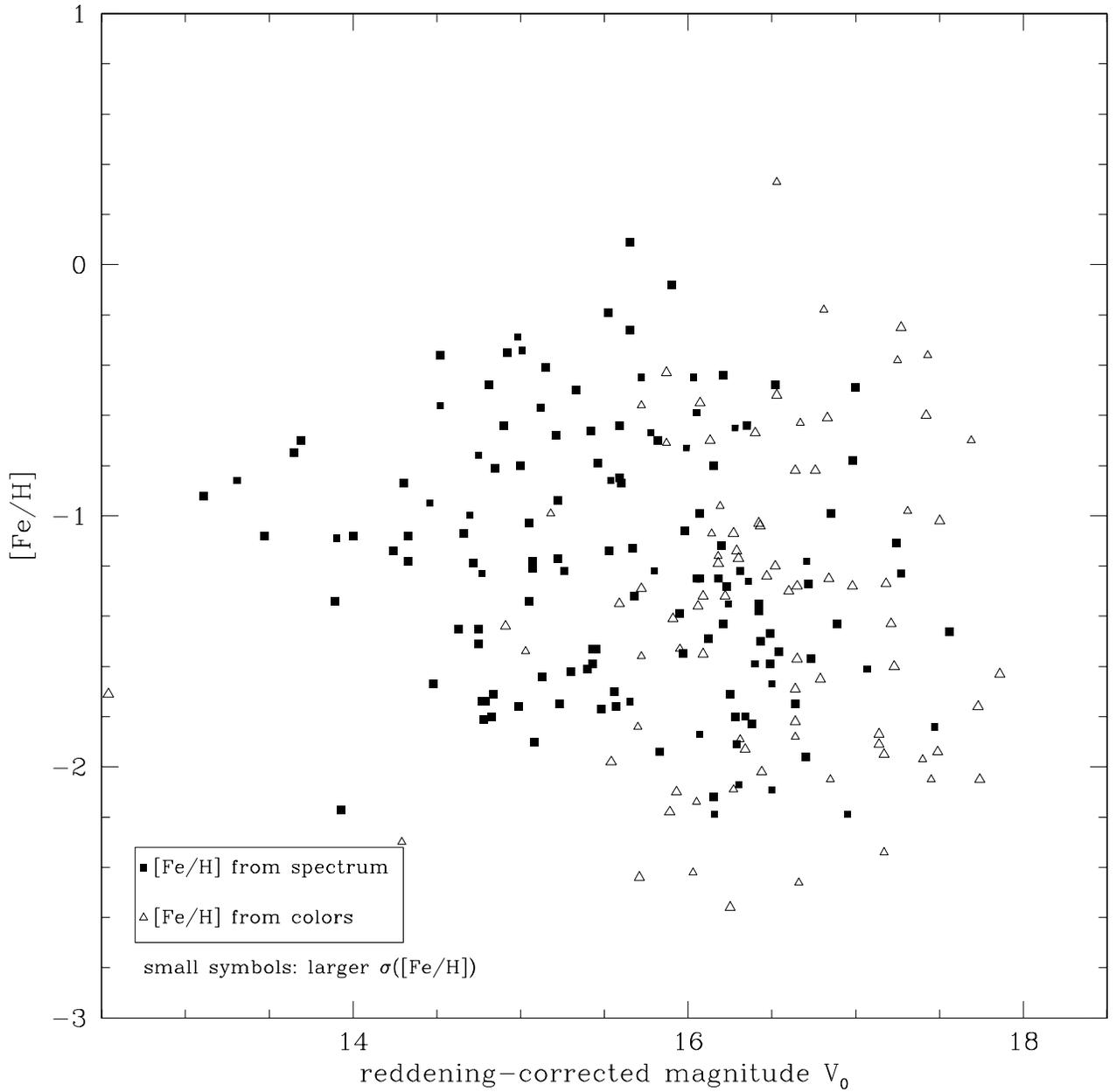}
\caption{[Fe/H] vs.\ $V_0$, dereddened total magnitude, for M31 GCs. The brightest object
is 037-B327; see discussion in Section~\ref{sec-red}.\label{lum_met}}
\end{figure}

The relation of cluster mass to metallicity is important in
globular cluster formation theory. If self-enrichment is important in
GCs, massive clusters should be more metal-rich; the opposite is
true if cooling from metals determines the temperature (and thus the
Jeans mass) in the cluster-forming clouds. Globular cluster 
destruction rates are higher for low-mass clusters closer to the galaxy center
\citep{az98}. If significant destruction has occurred in the
M31 GCS, one might expect there to be few faint, high-metallicity
clusters, since the highest-metallicity clusters are near the nucleus
and hence would have a greater chance of destruction.
In Figure~\ref{lum_met}
we plot metallicity versus (dereddened) apparent magnitude; as in
a similar plot by HBK, no trend is obvious. Least-squares fits both  
binned and unbinned in $V_0$ show no evidence for non-zero slopes,   
so we conclude that there is no evidence for a relationship between
luminosity (and presumably mass) and metallicity in the M31 clusters.

\section{Conclusions}
We have presented the results from a catalog of photometric
and spectroscopic information for M31 globular clusters.
We determine the reddening for 314 objects, with 221
of these values considered reliable. From the color excesses of
clusters with spectroscopic metallicities, we find that the
M31 and Galactic extinction laws are consistent.
The M31 and Galactic GC color-metallicity relations
are also consistent, and we use these relations to estimate
metallicities for M31 clusters without spectroscopic data.

The average intrinsic colors of M31 clusters are consistent
with those of Galactic clusters: the slightly higher mean metallicity
of the M31 clusters does not make a measurable difference
in their colors. There are no significant trends in M31 cluster
color with projected radius or distance along the disk axes.
The optical colors of M31 and Galactic GCs in two-color diagrams
agree fairly well with the predictions of population synthesis 
models after the models have been corrected for known defects.
However, there are significant ($0.05-0.2^m$) additional corrections required
for the predicted optical-infrared and infrared colors to match the data.
This indicates the presence of systematic
errors in the models, and the fact that corrections are required
so that the simplest model predictions -- broadband colors -- agree
with observations of the simplest stellar populations available
-- globular clusters -- is disturbing.
It is important to understand and remedy the
problems in the models before attempting to use them to 
study systems comprised of multiple populations.

The distributions of the most metal-sensitive colors, and of metallicity,
show evidence for bimodality. The two metallicity groups 
have means of [Fe/H]$\approx-1.4$ and $-0.6$. The metal-poor clusters
have a larger average projected distance from the galaxy, and show
slower rotation near the nucleus than the metal-rich clusters.
These properties suggest that the two metallicity groups are
analogs of the Galactic `halo' and `bulge/disk' clusters.
The presence of these two distinct populations in the
globular clusters as well as the stars emphasizes that GCS formation is
intimately related to galaxy formation.
The cluster system shows a small overall metallicity gradient, 
which implies that
the enrichment timescale for the proto-galactic gas was shorter than
the collapse timescale. There is no correlation between luminosity and metallicity,
which implies that neither self-enrichment or cooling from metals is
important in GC formation. The presence of faint, high-metallicity 
clusters in the galaxy disk constrains the destruction rate
of such objects. 
The M31 globular cluster system is very similar to the Galactic
system in many respects: mean metallicity, presence of 
two metallicity groups, broadband color distributions.
The possible presence of young globular clusters in M31
is an important difference between the properties of the two GCSs.
The properties of the two galaxies themselves also show similarities
and differences, and the relations between the galaxies and
their globular cluster systems remain important clues in the
study of galaxy formation. The detailed study of these two most
accessible systems of globular clusters provides an important
stepping stone on the path to understanding galaxy
and cluster system formation in more distant, younger, systems.

\acknowledgements

We thank the referee for helpful comments.
We thank A. Szentgyorgyi \& J. Geary for building the 4-Shooter camera, 
and S. Willner \& E. Tollestrup for building the SAO IR Camera.
We thank D. Koranyi, P. Garnavich, J. Mader, \& L. Macri
for help in acquiring observational data, and K. Stanek for helpful
conversations. We thank P. Hall, M. Akritas \& M. Bershady, and C. Bird,
respectively, for making the PHIIRS, BCES, and KMM software packages publicly available.
This research was supported by the Smithsonian
Institution, NATO Collaborative Research Grant CRG971552,
Faculty Research Funds from the University of California, Santa Cruz, 
and NSF grant AST-990732.

\clearpage

\begin{deluxetable}{lccccccccccl}
\tabletypesize{\scriptsize}
\tablewidth{0pt}
\tablenum{1}
\tablecaption{Catalog of photometric data for M31 globular clusters\label{oldphot}}
\tablehead{\colhead{name}&\colhead{$V$}&\colhead{\bv}&\colhead{\ub}&\colhead{\vr}&\colhead{$V-I$}&\colhead{$J$}&\colhead{$H$}&\colhead{$K$}&\colhead{opt source}&\colhead{IR source}& \colhead{comments}}
\startdata
000-001   & 13.75 &  0.83 & 0.29& 0.56&\nodata&11.84& 11.20& 11.04 &(1,1,2,1)& (3)            & CMD4,HRI1,HRS1    \\         
000-002   & 15.81 &  0.68 & 0.26& 0.48&\nodata&13.97& 13.53& 13.43 &(1,1,4,1)& (5) 	      &	HRS1     \\         
000-260   & 17.01 &  0.80 & 0.06& 0.48&\nodata&15.33& 14.55& 14.52 &(6,6,2,6)& (7) 	      &	HRI1     \\         
000-268   & 16.63 &  0.96 & 0.54& 0.62&\nodata&14.43& 13.82& 13.54 &(6,6,6,6)& (7) 	      &	\nodata    \\               
000-327   & 15.94 &  0.73 & 0.20& 0.54&\nodata&14.20& 13.69& 13.52 &(6,6,8,6)& (3) 	      &	CMD2,HRI1\\         
TRUNCATED\\
\enddata
\end{deluxetable}

\begin{deluxetable}{llc}
\tablewidth{0pt}
\tablenum{2}
\tablecaption{M31 cluster candidates shown not to be clusters\label{notclusters}}
\tablehead{\colhead{name}&\colhead{classification}&\colhead{reference}}
\startdata
000-003 &  galaxy+star & (1) \\
000-004 &  galaxy & (1) \\
000-005 &  galaxy & (1) \\
000-006 &  galaxy & (1) \\
TRUNCATED\\
\enddata
\end{deluxetable}

\clearpage
\begin{deluxetable}{lcccccccccccc}
\tablenum{3}
\tabletypesize{\tiny}
\rotate
\tablewidth{0pt}
\tablecaption{Photometric offsets for new optical photometry\label{optphotcomp}}
\tablehead{\colhead{}&\colhead{$\Delta V$}&\colhead{}&\colhead{}&\colhead{$\Delta(B\!-\!V)$}&\colhead{}&\colhead{}&\colhead{$\Delta(U\!-\!B)$}&\colhead{}&\colhead{}&\colhead{}&\colhead{}&\colhead{}\\
\colhead{}&\colhead{mean}&\colhead{$\sigma$}&\colhead{$n$}&\colhead{mean}&\colhead{$\sigma$}&\colhead{$n$}&\colhead{mean}&\colhead{$\sigma$}&\colhead{$n$}&\colhead{}&\colhead{}&\colhead{}}
\tablecolumns{13}
\startdata
\sidehead{PG}
all &  0.058$\pm$0.020 & 0.287&203&0.050$\pm$0.015& 0.205&173&\nodata&\nodata&\nodata\\
B87 &  0.029$\pm$0.029 & 0.297&104&0.098$\pm$0.023& 0.215&85& 0.035$\pm$ 0.028& 0.224&65\\ 
B82 & -0.032$\pm$0.015 & 0.150&92&-0.034$\pm$0.019& 0.048&88&\nodata&\nodata&\nodata\\
C85 &  0.741$\pm$0.211 & 0.560&7&\nodata&\nodata&\nodata&\nodata&\nodata&\nodata\\
\sidehead{PE}
all &0.037$\pm$0.010 & 0.134&188&0.024$\pm$0.010& 0.120&157&0.082$\pm$ 0.018& 0.199&123\\
SL83&0.037$\pm$0.014 & 0.137&98 &0.032$\pm$0.014& 0.132&91& 0.066$\pm$ 0.019& 0.161&68\\ 
SL85+&0.036$\pm$0.016& 0.123&70& 0.012$\pm$0.012& 0.100&66& 0.101$\pm$ 0.032& 0.238&55\\
\sidehead{(previous photometry: PG-PE)}
all &0.018$\pm$0.018 & 0.240&175&0.005$\pm$0.016& 0.215&175&0.003$\pm$ 0.031& 0.406&175\\
\sidehead{CCD}
        & $\Delta V$ & & & $\Delta(B\!-\!V)$ & & & $\Delta(V\!-\!R)$ & & & $\Delta(V\!-\!I)$\\
        &mean &$\sigma$ & $n$ & mean &$\sigma$ &$n$& mean &$\sigma$ & $n$ & mean &$\sigma$ & $n$\\
all &  0.022$\pm$  0.029 & 0.197&45& &\nodata&\nodata &\nodata &\nodata &\nodata& -0.125$\pm$ 0.042& 0.259&38\\
B93 &  0.053$\pm$  0.066 & 0.272&17 &  0.322$\pm$ 0.032& 0.125&15&-0.371$\pm$ 0.041& 0.163&16& -0.135$\pm$ 0.074& 0.307&17\\ 
R92 & -0.003$\pm$  0.019 & 0.046&6&   -0.034$\pm$ 0.019& 0.048&6&  0.065$\pm$ 0.041& 0.101&6 & \nodata          &\nodata&\\
M98 &  0.005$\pm$  0.032 & 0.151&22&  & & & & &  & -0.116$\pm$ 0.048& 0.220&21\\
\sidehead{internal scatter, this work}
\nodata& \nodata & 0.053&41 & \nodata & 0.076&41& \nodata & 0.082&41& \nodata & 0.082&41\\ 
\enddata
\tablecomments{Offsets are always in the sense (previous photometry)--(this work)}
\end{deluxetable}

\clearpage
\begin{deluxetable}{lcccccccl}
\tablewidth{0pt}
\tablenum{4}
\tablecaption{New photometric data for M31 globular clusters\label{newphot}}
\tablehead{\colhead{name}&\colhead{$V$}&\colhead{\bv}&\colhead{\ub}&\colhead{\vr}&\colhead{$V-I$}&\colhead{$J$}&\colhead{$K$} & \colhead{comments\tablenotemark{a}}}
\startdata
002-043 & 17.55(1)& 0.63(2) & -0.04(3)& 0.43(3) & 0.97(2)  & \nodata  & \nodata  & \nodata\\      
003-045 & 17.57(1)& 0.78(2) & 0.05(4) & 0.50(3) & 1.16(2)  & \nodata  & \nodata  & \nodata\\      
004-050 & 16.95(1)& 0.92(1) & 0.42(3) & 0.59(2) & 1.22(1)  & 14.91(2) & 14.19(5) & \nodata\\      
005-052 & 15.44(1)& 0.60(1) & 0.08(2) & 0.45(1) & 0.78(1)  & 14.16(2) & 13.81(6) & ID\\           
006-058 & 15.53(1)& 0.96(1) & 0.45(2) & 0.56(1) & 1.22(1)  & \nodata  & \nodata  & \nodata\\      
TRUNCATED\\
\enddata
\end{deluxetable}

\begin{deluxetable}{lrrl}
\tablenum{5}
\tablewidth{0pt}
\tablecaption{New spectroscopic data for  M31 globular clusters\label{newspectdata}}
\tablecolumns{4}
\tablehead{\colhead{}& \colhead{velocity}&\colhead{[Fe/H]}&\colhead{}\\
\colhead{name}& \colhead{(km~s$^{-1}$)} &\colhead{(dex)}&\colhead{comments}}
\startdata
025-084   & $-230\pm41$  & $-1.43\pm0.18$ & \nodata \\
036-000   & $-341\pm24$  & $-0.99\pm0.25$ & \nodata \\
125-183   & $-514\pm54$  & $-1.71\pm0.14$ & \nodata \\
126-184   & $-182\pm14$  & $-1.20\pm0.47$ & \nodata \\
134-190   & $-401\pm32$  & $-1.12\pm0.16$ & \nodata \\
TRUNCATED\\
\enddata
\end{deluxetable}

\clearpage
\begin{deluxetable}{ccc}
\tablenum{6}
\tablewidth{0pt}
\tablecolumns{3}
\tablecaption{Extinction law derived from M31 globular clusters\label{extrat}}
\tablehead{\colhead{} & \multicolumn{2}{c}{$E_{X-Y}/E_{B-V}$}\\
\colhead{$X-Y$} &\colhead{MW\tablenotemark{a}} & \colhead{M31}}
\startdata
\ub & 0.72 & $0.72\pm0.15$ \\ 
\uv & 1.72 & $1.54\pm0.14$ \\
$U-R$ & 2.30 & $2.19\pm0.22$ \\
\br & 1.58 & $1.61\pm0.16$ \\ 
\bi & 2.26 & $2.45\pm0.23$ \\
\vi & 1.26 & $1.40\pm0.22$ \\
\vk & 2.75 & $2.48\pm0.33$ \\
\jk & 0.52 & $0.53\pm0.19$ \\
\enddata
\tablenotetext{a}{From \citet{ccm89}.}
\end{deluxetable}

\begin{deluxetable}{lcccccccc}
\tablenum{7}
\tablewidth{0pt}
\tablecaption{Color-metallicity relations for Galactic GCs\label{color-met}}
\tablecolumns{6}
\tablehead{\colhead{}&\multicolumn{2}{l}{$(X-Y)_0= a{\rm [Fe/H]}+ b$}&\multicolumn{2}{l}{${\rm [Fe/H]}=a(X-Y)_0+ b$}&\colhead{}\\
\colhead{$(X-Y)_0$}&\colhead{$a$}&\colhead{$b$}&\colhead{$a$}&\colhead{$b$}&\colhead{N}}
\startdata
$(B-V)_0$ &$0.159\pm0.011$& $0.92\pm0.02$ & $ 5.50\pm0.33$ & $-5.26\pm0.23$	&  88\\ 
$(B-R)_0$ &$0.262\pm0.014$& $1.51\pm0.02$ & $ 3.69\pm0.26$ & $-5.62\pm0.30$	&  66\\ 
$(B-I)_0$ &$0.318\pm0.024$& $2.07\pm0.04$ & $ 2.79\pm0.26$ & $-5.94\pm0.42$	&  76\\ 
$(U-B)_0$ &$0.289\pm0.018$& $0.57\pm0.03$ & $ 2.76\pm0.20$ & $-1.86\pm0.04$	&  81\\ 
$(U-V)_0$ &$0.457\pm0.026$& $1.50\pm0.04$ & $ 1.92\pm0.09$ & $-3.05\pm0.09$	&  81\\ 
$(U-R)_0$ &$0.572\pm0.027$& $2.11\pm0.04$ & $ 1.62\pm0.09$ & $-3.52\pm0.11$	&  66\\ 
$(V-I)_0$ &$0.156\pm0.015$& $1.15\pm0.02$ & $ 4.22\pm0.39$ & $-5.39\pm0.35$	&  75\\ 
$(J-K)_0$ &$0.177\pm0.021$& $0.91\pm0.03$ & $ 5.86\pm0.86$ & $-5.25\pm0.52$	&  37\\ 
$(V-K)_0$\tablenotemark{a}&$0.593\pm0.080$& $3.15\pm0.12$ & $ 1.30\pm0.15$ & $-4.45\pm0.36$&  23\\ 
$(V-K)_0$\tablenotemark{b}&$0.611\pm0.070$& $3.14\pm0.10$ & $ 1.40\pm0.17$ & $-4.62\pm0.38$&  35\\ 
\enddata
\tablenotetext{a}{low $E_{B-V}$ only}
\tablenotetext{b}{all data}
\end{deluxetable}

\clearpage

\begin{deluxetable}{lcccccc}
\tablenum{8}
\tablecolumns{7}
\tablewidth{0pt}
\tablecaption{Distribution of intrinsic colors for M31 clusters\label{tab-colordist}}
\tablehead{\colhead{color}&\colhead{M31}&\colhead{}&\colhead{}&\colhead{MW}&\multicolumn{2}{l}{M31 predicted}\\
\colhead{}&\colhead{mean}&\colhead{median}&\colhead{$\sigma$}&\colhead{mean}&\colhead{metal-poor}&\colhead{metal-rich}}
\startdata
$(B-V)_0$ &   $0.72 \pm0.01$ & 0.72 & 0.12 &$0.71\pm0.01$ & 0.68 & 0.83 \\	
$(B-R)_0$ &   $1.18 \pm0.01$ & 1.19 & 0.12 &$1.18\pm0.02$ & 1.12 & 1.35 \\	
$(B-I)_0$ &   $1.68 \pm0.01$ & 1.68 & 0.16 &$1.64\pm0.02$ & 1.59 & 1.88 \\	
$(U-B)_0$ &   $0.16 \pm0.02$ & 0.16 & 0.23 &$0.20\pm0.02$ & 0.13 & 0.39 \\	
$(U-V)_0$ &   $0.88 \pm0.02$ & 0.84 & 0.31 &$0.89\pm0.03$ & 0.81 & 1.23 \\	
$(U-R)_0$ &   $1.35 \pm0.03$ & 1.32 & 0.34 &$1.38\pm0.03$ & 1.25 & 1.77 \\	
$(V-R)_0$ &   $0.46 \pm0.01$ & 0.47 & 0.05 &$0.47\pm0.01$ & 0.45 & 0.52 \\	
$(V-I)_0$ &   $0.96 \pm0.01$ & 0.96 & 0.11 &$0.94\pm0.01$ & 0.92 & 1.06 \\	
$(J-K)_0$ &   $0.67 \pm0.01$ & 0.68 & 0.13 &$0.64\pm0.02$ & 0.64 & 0.81 \\	
$(V-K)_0$ &   $2.32 \pm0.02$ & 2.32 & 0.26 &$2.24\pm0.05$ & 2.26 & 2.72 \\	
\enddata
\end{deluxetable}

\begin{deluxetable}{lccccccccc}
\tablenum{9}
\tablewidth{0pt}
\tablecaption{Distribution of [Fe/H] for M31 clusters\label{fehdist}}
\tablehead{\colhead{}&\colhead{ }&\colhead{ }&\colhead{ }&\colhead{KMM}\\
\colhead{dataset}&\colhead{$\overline{\rm [Fe/H]}$}&\colhead{${\sigma}_{\rm [Fe/H]}$}&\colhead{median [Fe/H]}&\colhead{$\overline{\rm [Fe/H]}_1$}&\colhead{$\overline{\rm [Fe/H]}_2$}&\colhead{$n_1$}&\colhead{$n_2$}&\colhead{${\sigma}_{\rm [Fe/H]}$}&\colhead{$p$}}
\startdata
1  & -1.22 $\pm$ 0.04 & 0.58 &	-1.25 &	-1.48 &	-0.63 &	169 &  78 &   0.43  &  0.102\\
1a & -1.21 $\pm$ 0.04 & 0.53 &	-1.25 &	-1.42 &	-0.64 &	135 &  43 &   0.40  &  0.098\\
2  & -1.15 $\pm$ 0.04 & 0.54 &	-1.23 &	-1.43 &	-0.60 &	110 &  56 &   0.38  &  0.042\\
2a & -1.14 $\pm$ 0.05 & 0.52 &	-1.19 &	-1.36 &	-0.53 &	94  &  31 &   0.37  &  0.034\\
\enddata
\end{deluxetable}

\clearpage

\begin{deluxetable}{lcccccc}
\tablenum{10}
\tablewidth{0pt}
\tablecaption{Comparison of spiral galaxy globular cluster systems\label{gcscomp}}
\tablehead{\colhead{galaxy}&\colhead{$N_{GC}$}&\colhead{$M_V$}&\colhead{$S_N$}&\colhead{[Fe/H]}&\colhead{${\sigma}_v$}&\colhead{Ref.}} 
\startdata
MW total   &$180\pm20$   &$-21.3$&$0.5\pm0.1$ & $-1.34\pm0.07$ &\nodata&(1)\\
MW halo	   &0.69         &\nodata&\nodata&-1.59 &100  & (2) \\
MW disk	   &0.31         &\nodata&\nodata&-0.55 &124  & (2) \\
M31 total  &$450\pm100$	 &$-21.8$ &$0.9\pm0.2$ &$-1.15\pm0.04$&142  & (1,3) \\
M31 halo   &0.66         &\nodata&\nodata&-1.43 &148  & (3) \\
M31 disk   &0.34         &\nodata&\nodata&-0.60 &146  & (3) \\
M33	   &$\sim25$	 &$-19.4$  &$0.6\pm0.2$ &-1.6 &70   & (1,4) \\
M81	   &$210\pm30$	 &$-21.1$  &$0.7\pm0.1 $&$-1.48\pm0.19$&152 & (5,6) \\
M104       &$1600\pm800$ &$-22.2$  &$2\pm1$     &$-0.7\pm0.3$ &260 & (7,8) \\
\enddata
\tablecomments{Velocity dispersions ${\sigma}_v$ are in km~s$^{-1}$. Numbers in $N_{GC}$
column for M31/MW disk/halo are fraction of total.}
\tablerefs{(1) \citealt{az98}; (2) \citealt{cot99}; (3) this work; 
(4) \citealt{sch93}; (5) \citealt{pr95}; (6) \citealt{pbh95}; (7) \citealt{bh92};
(8) \citealt{bri97}}
\end{deluxetable}

\clearpage

\clearpage

\clearpage

\clearpage

\clearpage

\clearpage

\clearpage

\clearpage

\clearpage

\clearpage

\clearpage

\clearpage

\clearpage

\clearpage

\clearpage

\clearpage

\clearpage

\clearpage

\clearpage 

\clearpage

\end{document}